\PassOptionsToPackage{unicode}{hyperref}
\PassOptionsToPackage{hyphens}{url}
\documentclass[
]{article}
\usepackage{lmodern}
\usepackage{amssymb,amsmath}
\usepackage{ifxetex,ifluatex}
\ifnum 0\ifxetex 1\fi\ifluatex 1\fi=0 
  \usepackage[T1]{fontenc}
  \usepackage[utf8]{inputenc}
  \usepackage{textcomp} 
\else 
  \usepackage{unicode-math}
  \defaultfontfeatures{Scale=MatchLowercase}
  \defaultfontfeatures[\rmfamily]{Ligatures=TeX,Scale=1}
\fi
\IfFileExists{upquote.sty}{\usepackage{upquote}}{}
\IfFileExists{microtype.sty}{
  \usepackage[]{microtype}
  \UseMicrotypeSet[protrusion]{basicmath} 
}{}
\makeatletter
\@ifundefined{KOMAClassName}{
  \IfFileExists{parskip.sty}{%
    \usepackage{parskip}
  }{
    \setlength{\parindent}{0pt}
    \setlength{\parskip}{6pt plus 2pt minus 1pt}}
}{
  \KOMAoptions{parskip=half}}
\makeatother
\usepackage{xcolor}
\IfFileExists{xurl.sty}{\usepackage{xurl}}{} 
\IfFileExists{bookmark.sty}{\usepackage{bookmark}}{\usepackage{hyperref}}
\hypersetup{
  pdftitle={Differential analysis in Transcriptomic},
  pdfauthor={Dorota Desaulle, Céline Hoffmann, Bernard Hainque and Yves Rozenholc},
  hidelinks,
  pdfcreator={LaTeX via pandoc}}
\usepackage[margin=1in]{geometry}
\usepackage{graphicx,grffile}
\makeatletter
\def\maxwidth{\ifdim\Gin@nat@width>\linewidth\linewidth\else\Gin@nat@width\fi}
\def\maxheight{\ifdim\Gin@nat@height>\textheight\textheight\else\Gin@nat@height\fi}
\makeatother
\setkeys{Gin}{width=\maxwidth,height=\maxheight,keepaspectratio}
\makeatletter
\def\fps@figure{htbp}
\makeatother
\usepackage{graphicx}
\usepackage{color}
\usepackage{xcolor}
\usepackage{geometry}
\usepackage{hyperref}
\usepackage{amsmath,amssymb,MnSymbol,marvosym}
\usepackage{multirow}
\usepackage{booktabs}
\usepackage{float,tabls,array}
\usepackage[T1]{fontenc}
\usepackage[official,right]{eurosym}
\usepackage{fancybox}
\usepackage{ulem}
\usepackage[french]{babel}
\usepackage{times}
\usepackage{fancyhdr}
\usepackage{bbm}
\usepackage{ulem}
\usepackage{dashrule}
\usepackage{enumerate}
\usepackage{afterpage}
\def\begincols{\begin{columns}}
\def\begincol{\begin{column}}
\def\endcol{\end{column}}
\def\endcols{\end{columns}}
\definecolor{DDgreen}{RGB}{0,190,0}
\definecolor{latexBleu}{RGB}{50,50,180}

\usepackage{bbm}
\usepackage{setspace}
\let\rmarkdownfootnote\footnote%
\def\footnote{\protect\rmarkdownfootnote}

\usepackage{titling}
\pretitle{\vspace{\droptitle}\centering\huge}
\posttitle{\par}

\preauthor{\centering\large\emph}
\postauthor{\par}

\predate{\centering\large\emph}
\postdate{\par}

\title{Differential analysis in Transcriptomic}
\usepackage{etoolbox}
\makeatletter
\providecommand{\subtitle}[1]{
  \apptocmd{\@title}{\par {\large #1 \par}}{}{}
}
\makeatother
\subtitle{The strength of randomly picking `reference' genes}
\author{Dorota Desaulle, Céline Hoffmann, Bernard Hainque and Yves Rozenholc}
\date{Version 23 mars, 2021}

\begin{document}
\maketitle

\newcommand\cs{\mathrel{\overset{\makebox[0pt]{\mbox{\normalfont\tiny\sffamily CS}}}{\leq}}}
\newcommand{\A}{A}
\newcommand{\B}{B}
\newcommand{\diag}{\textrm{diag}}
\newcommand{\tr}{\textrm{Tr}}
\newcommand{\rk}{\textrm{rank}}
\newcommand{\smax}{s_{\max}}
\newcommand{\smin}{s_{\min}}
\newcommand{\pr}{{P^{\mathcal R}}}
\newcommand{\cov}{\textrm{cov}}
\newcommand{\corr}{\textrm{corr}}
\newcommand{\var}{\textrm{var}}
\newcommand{\red}[1]{\textcolor{red}{#1}}
\newcommand{\blue}[1]{\textcolor{blue}{#1}}
\newcommand{\NB}[1]{\textcolor{violet}{#1}}
\newcommand{\DDgreen}[1]{\textcolor{DDgreen}{#1}}
\newtheorem{assumption}{Assumption}
\newtheorem{theo}{Theorem}
\newtheorem{proc}{Procedure}
\newtheorem{lem}{Lemma}
\newtheorem{cor}{Corrolary}
\newtheorem{prop}{Property}
\newtheorem{definition}{Definition}

\paragraph*{Affiliations}

D.D. and Y.R. : Université de Paris, UR 7537 - BioSTM, Biostatistique,
Traitement et Modélisation des données biologiques, F-75006 Paris,
France C.H. and B.H. : Université de Paris, CNRS, INSERM - UTCBS, Unité
des technologies Chimiques et Biologiques pour la Santé, F-75006 Paris,
France

\paragraph*{Contributions}

Y.R. conceived the iterative use of randomly picked genes to derive a
proper differential analysis free of any knowledge of so-called
housekeeping genes. D.D. and Y.R. did the mathematical modeling and
analysis of the procedure. D.D. and Y.R. implemented the procedure to
confirm empirically this analysis and derive supplementary results about
the power. They performed the differential analysis on the real data.

C.H. and B.H. conceived and designed the biological experiments. C.H.
performed the biological experiments. C.H. and B.H. commented the
results obtained on real data.

\paragraph*{Corresponding author}

Y.R. \texttt{\textless{}yves.rozenholc@u-paris.fr\textgreater{}}

\hypertarget{abstract}{%
\section*{Abstract}\label{abstract}}
\addcontentsline{toc}{section}{Abstract}

Transcriptomic analysis are characterized by being not directly
quantitative and only providing relative measurements of expression
levels up to an unknown individual scaling factor. This difficulty is
enhanced for differential expression analysis. Several methods have been
proposed to circumvent this lack of knowledge by estimating the unknown
individual scaling factors however, even the most used one, are
suffering from being built on hardly justifiable biological hypotheses
or from having weak statistical background. Only two methods withstand
this analysis: one based on largest connected graph component hardly
usable for large amount of expressions like in NGS, the second based on
\(\log\)-linear fits which unfortunately require a first step which uses
one of the methods described before.

We introduce a new procedure for differential analysis in the context of
transcriptomic data. It is the result of pooling together several
differential analyses each based on randomly picked genes used as
reference genes. It provides a differential analysis free from the
estimation of the individual scaling factors or any other knowledge.
Theoretical properties are investigated both in term of FWER and power.
Moreover in the context of Poisson or negative binomial modelization of
the transcriptomic expressions, we derived a test with non asymptotic
control of its bounds. We complete our study by some empirical
simulations and apply our procedure to a real data set of hepatic miRNA
expressions from a mouse model of non-alcoholic steatohepatitis (NASH),
the CDAHFD model. This study on real data provides new hits with good
biological explanations.

\hypertarget{introduction}{%
\section{Introduction}\label{introduction}}

Transcriptomic analysis of a tissue sample (an individual) results in
the measurements inside either a fix volume or a fix weight of
expressions of some given set of expressions (that for simplicity we
will consider gene expressions). As the amount of molecules inside the
fix analyzed quantity is not controlled, it is known that all these
measurement are scaled by a single unknown factor depending of the
individual at hand. In other words, transcriptomic analysis are
characterized by being not directly quantitative and by only providing
relative measurements of expression levels up to an unknown individual
scaling factor. When some of this genes are housekeeeping genes with
known expression properties, they serve as reference genes and one can
use their observed relative expression levels to get a normalization
(Vandesompele et al.
\protect\hyperlink{ref-vandesompele_accurate_2002}{2002}). However, in
exploratory differential analysis, such reference genes cannot always be
known in advance and we therefore have to compare several sets of
expression levels, each set depending of the unknown scaling factor
attached to the corresponding individual.

Apart from the crude normalization by the \emph{total count} (Marioni et
al. \protect\hyperlink{ref-marioni_rna-seq:_2008}{2008}; Mortazavi et
al. \protect\hyperlink{ref-mortazavi_mapping_2008}{2008}), several
methods have been proposed to circumvent this issue: \emph{upper
quantile} (Bullard et al.
\protect\hyperlink{ref-bullard_evaluation_2010}{2010}), \emph{trimmed
mean of M values} (TMM) (Mark D. Robinson and Oshlack
\protect\hyperlink{ref-robinson_scaling_2010}{2010}) and
\emph{interindividual median count ratio accross gene} (Anders and Huber
\protect\hyperlink{ref-anders_differential_2010}{2010}), which can be
found in the Bioconductor packages \emph{DESeq2} (Love, Huber, and
Anders \protect\hyperlink{ref-DESeq2}{2014}) and \emph{EdgeR} (Mark D
Robinson, McCarthy, and Smyth \protect\hyperlink{ref-edgeR}{2010}). All
these methods are based on the belief that reference genes may be
identified as their expressions are expected to be less variable in the
overall population and hence less variable even in presence of
fluctuations of the scaling factors. This belief is neither proven nor
mathematically justifiable. Counter-examples can be built for example by
considering reference genes showing more variability than non-reference
ones.

More recently, Li et al.
(\protect\hyperlink{ref-li_normalization_2012}{2012}) proposed to use
\(\log\)-linear fits to detect DE genes, however it also relies on a
scaling factor estimation achieved by starting from the total count to
selected iteratively a subset of genes associated with small values of a
Poisson goodness-of-fit statistic.

Up to our knowledge, only Curis et al.
(\protect\hyperlink{ref-curis_determination_2019}{2019}), have proposed
an approach free of this preliminary selection of reference genes and of
the estimation of the scaling factors. Having in mind that the ratio of
two expressions for one individual is free from the unknown scaling
factor, only expression ratios are compared. To this end, they are
considered to be vertices of a graph. Two vertices are connected if the
expression ratios may be considered as equal between the two
populations. The largest connected component of the graph is expected to
be made of non differential genes and DE genes are expected to live
outside of this largest component. Without discussing the hypotheses
used to build the largest connected component, it is easily
understandable that the number of vertices of this graph being the
square of the number of expressions, such an approach is reserved to
transcriptomic studies like qPCR where the number of expressions is
small, unlike high-throuput experiments which are our interest.

In brief, actual procedures for differential analysis in such
high-throuput transcriptomic experiments are build on a preliminary
step, which consists in finding some non differential expressions to
estimate the scaling factors. Then data are reused for testing. It is
not only unsatisfactory to lack a good recipe for this first step, but
also unproper and statistically worst, to do a differential analysis by
having to run at first a non-differential analysis on the same data.

In view of this drawback that affects the methods in use, our proposal
does not aim at finding first some non differential expressions to
rescale the data. Instead, we propose an iterative framework in which a
rescaling is realized at each step of the iteration from a randomly
selected subset of expressions which are no longer assumed to be non
differentially expressed, the differential analysis being performed on
the remaining data. In this sense, our proposal relies only on the use
of a differential test, which is only expected to have good type I and
type II errors when the data are well scaled, without an estimation of
the ``real'' scaling factors, which may be estimated afterwards.

In this transcriptomic context, for a given individual \(i\) and a gene
\(j\), the measure is modelized by a Poisson random variable \(X_{ij}\)
with intensities \(\lambda_{ij}\) being the product of an individual
scaling factor \(s_i\), which only depends on the individual \(i\),
times a gene dependent expression level \(\mu_j^A\) or \(\mu_j^B\) with
the upper script refering to the conditions \(A\) or \(B\). To account
for observed over-dispersion with respect to the Poisson model, gamma
convoluted modelization of this model is also used leading to the
so-called negative binomial model. This latter modelization is all the
more suitable when transcriptomic expressions are measured at large
genomic scale like gene expression levels pooling together reads of
non-homogeneous origins over few hundreds of base-pairs, unlike miRNA or
siRNA experiments which cover around 100 base-pairs and are more likely
to be mimicked by simple Poisson random variables. In both
modelizations, the expectation satisfies
\[\mathbb E(X_{ij}) = \lambda_{ij} = s_i \times \mu_j\] with \(\mu_j\)
representing \(\mu_j^A\) or \(\mu_j^B\).

In what follows, the expressions will be considered to be
gene-expressions, however they could come from any transcriptomic
experiment (RNA-seq, miRNA-seq, etc). The individuals belong to two
populations or have been studied under two differents conditions \(A\)
and \(B\), providing \(n_A\) and \(n_B\) individuals under each
condition. We aim to find those genes which are differentially expressed
(DE) genes between the two populations. We point out that transcriptomic
studies using miRNA or siRNA are often explanatory studies made on two
sub-populations of mice for example, such that they are characterized by
small sample sizes, \(n_A\) and \(n_B\) being usually between 5 and 10
individuals for each subpopulation.

Our intensive iterative random procedure for detection can be summarized
in more details as follows. At each step of the iteration, a random
subset of genes is selected and considered to be made of reference
genes, used to get a normalization. After this normalization, the
non-selected genes are tested for differential behaviors. Along the
iterations, the detections for each gene are pooled. After the
iterations, the pooled detections are compared to the rates of potential
wrong detections due to miss-picking randomly genes in the unknown set
of DE genes. Our method controls the FWER for any test procedure having
its level and power controled when the scaling factors are known. It is
adaptive to the unknown number of genes which would be detectable, given
the observations, if the scaling factors were known, assuming only that
the number of DE genes is less than half of the total number of genes.

Moreover, taking advantage that our procedure behaves as if reference
genes were available, we propose and study a unified testing procedure
for differential analysis, adapted to our random detector for the two
classical modelizations (Poisson and Negative binomial). This test
derives from a procedure where scaling factors would be known and in
this sense satisfies the requirements in term of type I and II errors of
our random procedure. Assuming that the expressions levels are high
enough, we study their properties. It is shown to be approximately a
standard Gaussian and we derive non-asymptotic control for this
approximation so that the test can have its level well controled at
finite distance.

To complete our study, we run an intensive simulation. Finally we apply
our procedure on a real data set of hepatic miRNA expressions from a
mouse model of non-alcoholic steatohepatitis (NASH), the CDAHFD
(choline-deficient, L-amino acid-defined, high-fat diet) model, with 4
cases of NASH with hepatic fibrosis and 4 controls without hepatic
lesions.

\hypertarget{formulation-of-the-problem}{%
\section{Formulation of the problem}\label{formulation-of-the-problem}}

We observe counts \(X_{ij}\) for \(i=1,...,n\) and \(j=1,...,m\) the
\(m\) expressions of \(n\) individuals belonging to two populations
caracterized by index subsets \(A\) and \(B\) of sizes \(n_A\) and
\(n_B\), with \(n=n_A+n_B\). The first \(n_A\) individuals belonging to
groupe \(A\).

In an homogeneous population, the expectation
\(\lambda_{ij}:=\mathbb E(X_{ij})\) is assumed to be the product of a
scaling factor \(s_i\) attached to the individual \(i\) together with
\(\mu_j\) an unknown relative quantification of gene \(j\) potentially
varying with the population, \(\lambda_{ij} = s_i\mu_j\). By convention,
we set \(\sum_{i=1}^n s_i=n\). With this convention, when all
experiments perform the same and \(A\) and \(B\) are two independent
sub-populations of the same population then \(s_i=1\) for
\(i=1,\ldots,n\). In the latter case, \(\mu_j\) stands for the relative
expression level of gene \(j\) in this population.

When two sub-populations \(A\) and \(B\) are considered, \(\mu_j\) is
assumed to depend only on the experimental condition under which \(i\)
is considered. In other words, \(\mu_j=\mu_j^A\) if \(i\in A\) and
\(\mu_j=\mu_j^B\) if \(i\in B\).\medskip 

For a given \(j\) in \(\{1,...,m\}\), we want to test whether the
\(j\)-th expression is differentially expressed under experimental
conditions \(A\) and \(B\). This can be stated as a test of the
following hypotheses \begin{equation}\label{eq:test-hyp-0}
H_0^j : \text{gene $j$ is not DE} \quad \text{ against } \quad H_1^j : \text{gene $j$ is DE}.
\end{equation}

In other words, distinguishing the relative quantification of gene \(j\)
in populations \(A\) and \(B\):\\
\begin{equation}\mathbb E(X_{ij})=s_i\, \mu_j^A,\,\text{ for } i\in A \qquad\text{ and }\qquad \mathbb E(X_{ij})=s_i\, \mu_j^B\,\text{ for } i\in B, \end{equation}
the hypotheses defined by \eqref{eq:test-hyp-0}, may be formulated as\\
\begin{equation}\label{eq:test-hyp-1}
H_0^j : \mu_j^A=\mu_j^B \quad \text{ against } \quad H_1^j : \mu_j^A\not = \mu_j^B.
\end{equation}

A gene \(j\) satisfying \(H_0^j\) is called ``invariant'' (between
conditions \(A\) and \(B\)) or a ``reference gene''.\medskip 

\begin{definition}[DE $\eta$-detectable] Let $h$ be an increasing function and the $s_i$, $i=1,...,n$ be \underline{known}. Suppose that we have at hand a symmetric test statistic $T^{*}$ with distribution only depending on the parameter $h(\mu^A_j) - h(\mu^B_j)$ and which is stochastically increasing with respect to this parameter. Let us denote by $q^{*}(1-\eta/2)$ an upper quantile of $T^{*}(X_{1 j},...,X_{n j})$ when  $\mu^A_{j}=\mu^B_{j}$. We say that gene $j$ is "DE $\eta$-detectable" at level $\eta>0$ if and only if the null hypothesis $H_0^j$ is rejected at this level that is if and only if 
\begin{equation}\label{eq:rejection-case-known}
|T^{*}(X_{1 j},...,X_{n j})|\geq q^{*}(1-\frac{\eta}{2}).
\end{equation}
The index set of the DE $\eta$-detectable genes is denoted $\mathcal D^*$.
\end{definition}

Clearly this definition depends, for all \(j\), on the observed sample
\((X_{1 j},...,X_{n j})\), hence all what follows including our test
construction and its analysis is conditional to this
observation.\medskip

The set \(\mathcal D^*\) may be empty but it is unlikely to be large,
unless the design of the experiment is irrelevant. By construction its
cardinal satisfies \(|\,\mathcal D^*\,|\,=d^*\) and we assume further
that \(d^*\) is samller than \(m/2\). \bigskip

Note that the quantile \(q^{*}(1-\eta/2)\) does not depend on parameters
\(\mu^A_{j}\) and \(\mu^B_{j}\) as \(T^{*}\) is assumed pivotal. The
symmetry of \(T^{*}\) is only here for simplicity. Clearly being DE
\(\eta\)-detectable depends on the choice of \(T^{*}\). A bad choice
could even conduct to declare all genes to be either not DE or DE.
\medskip

In the special case where the \(s_i\), \(i=1,...,n\) are known, we
denote by \(p_j^{*}\) the \(p\)-value associated to the test of the gene
\(j\) and the sorted \(p\)-values by \(p_{(j)}^{*}\) such that
\(p_{(j)}^{*}\leq p_{(j+1)}^{*}\). In order to control the family-wise
error rate (FWER) and account for multiple testing, several procedures
can be applied, we focus on the Holm's step-down procedure (Holm
\protect\hyperlink{ref-holm_1979}{1979}).

\begin{proc}[Holm's procedure] The $s_i$, $i=1,...,n$ being \underline{known}, let $d^{*}$ be the minimal index $j\geq 0$  such that $p_{(j+1)}^{*} \geq \alpha\big/(m-j)$, the genes corresponding to $p_{(1)}^{*}$, ..., $p_{(d^{*})}^{*}$ are declared differentially expressed with the convention that if $d^{*}=0$ then all gene are declared invariant. \end{proc}

This procedure is known to ensure that the FWER is less than \(\alpha\)

\paragraph*{Guide of lecture}

The next section introduces several notations. Section \ref{sec:testing}
is devoted to the construction of a practical statistic based on
randomly picked ``reference'' genes when the \(s_i\) are unknown which
is derived from \(T^{*}\). Section \ref{sec:our-proc} contains Theorem
\ref{th:random-picking} which proves that this construction controls the
FWER as soon as \(T^{*}\) satisfies simple conditions on the type I and
II errors. Section \ref{sec:power} presents Theorem \ref{th:fwer-power}
and its corollary which give necessary conditions on the number of
iterations to get FWER and exponential control of the power. Section
\ref{sec:testing-proc} focuses on the construction of a satisfying test
\(T^{*}\) when the distribution of gene expressions is assumed to follow
Poisson or Negative Binomial distribution. The adjusted rejection region
corresponding to \ref{eq:rejection-case-known} for a practical
implementation is a consequence of Theorem \ref{th:control} and its
corollary. In Section \ref{sec:empirical}, we study empirically the
behavior of our procedure including its level and its power. Section
\ref{sec:real-data} is devoted to the study of a real data set of
hepatic miRNA expressions from a mouse model. The proofs of our two
theorems are reported in Appendix.

\hypertarget{notations}{%
\section{Notations}\label{notations}}

\label{sec:notations}

We will further use the convention that \(\bullet\) in exponent
indicates the set --\(A\) or \(B\)-- the index \(i\) is belonging to.
For example \(\mu_j^\bullet\) denotes \(\mu_j^A\) if \(i\in A\) and
\(\mu_j^B\) if \(i\in B\). \medskip

Given two integers \(k\) and \(r\), we consider the sampling of \(r\)
subsets of indexes in \(\{1,\ldots,m\}\) of size \(k\), denoted
\(S_1,\ldots,S_r\). We denote by \(\mathcal S:=\{S_1,\ldots,S_r\}\).
Each subset from \(\mathcal S\) will be used to provide an estimation of
the unknown \(s_i\) and later refered as a \emph{normalisation subset}.
Given \(S\in \mathcal S\), the genes with index in \(S\) are used for
the estimation of the scaling factors \(s_i\). The remaining expressions
with index in \(\bar S := \{1,\ldots,m\} \setminus S\) are tested for
differential behavior between the two populations. \medskip

For a given \(j\) in \(\{1,\ldots,m\}\), we denote by
\(\mathcal S_j := \{S\in\mathcal S \,|\, j\not\in S\}\) the set of
subsets in \(\mathcal S\) which do not contain \(j\). The total number
of tests run for gene \(j\) is \(r_j:=|\mathcal S_j|\). Clearly
\(r_j\sim\mathcal B(r,\kappa)\) with \begin{equation}\label{eq:kappa}
\kappa :=P(S\in\mathcal S_j)=1-\binom{m-1}k\big/\binom m k=\frac{m-k}m=1-k/m.
\end{equation}

If a subset \(S_j\) contains at least one gene \(j\) in \(\mathcal D^*\)
then the normalization may be wrong. We call such subset a \emph{badly
selected subset} or a \emph{wrong normalization subset}. Note that a
such subset exists if \(\mathcal D^*\) is not empty, that is if
\(d^*>0\).

The number of wrong normalization subsets not containing \(j\) is
\begin{equation}\label{eq:Bj} B_j := \sum_{S\in\mathcal S_j} (S\cap\mathcal D^*\not=\emptyset) = |\{S\in\mathcal S_j,S\cap\mathcal D^*\not=\emptyset\}|.\end{equation}
Taking into account that \(j\) is DE \(\eta\)-detectable
(\(j\in\mathcal D^*\)) or not and assuming that \(d^*=d\), the random
variable \(B_j\) given the observation of \(r_j\) follows a binomial
distribution
\[B_j|r_j\, \sim\,\left\{\begin{array}{ccl}\mathcal B(r_j,\pi_d^0)&\text{ if }&j\not\in\mathcal D^*\medskip\\\mathcal B(r_j,\pi_d^1)&\text{ if }&j\in\mathcal D^*\end{array}\right.,\]
where \begin{equation}\label{eq:pid}
\pi_d^0:=\left\{\begin{array}{lcl} 1-{\binom{m-d-1}{k}}\Big/{\binom{m-1}{k}}   
& \text{ if} & 0\leq d \leq m-k-1\\
1 & \text{ if} & d \geq m-k \end{array}\right.\quad\text{ and }\quad\pi_d^1:=\left\{\begin{array}{lcl} 1-{\binom{m-d}{k}}\Big/{\binom{m-1}{k}}   
& \text{ if} & 0< d \leq m-k\\
1 & \text{ if} & d \geq m-k+1 \end{array}\right.
\end{equation} with the convention that \(\pi_0^1=0\). The number
\(\pi_d^0\) (respectively \(\pi_d^1\)) represents the probability that a
wrong normalization subset does not contain \(j\) when the latter is
invariant, (respectively DE \(\eta\)-detectable). \medskip

\begin{prop} The numbers $\pi_d^0$  and $\pi_d^1$ satisfy $\pi_d^0>\pi_d^1$ and the sequence $\pi_d^0\big/\pi_d^1$ is decreasing and converges to 1 when $d$ grows.\end{prop}

\hypertarget{testing-using-randomly-picking-reference-genes}{%
\section{\texorpdfstring{Testing using randomly picking reference genes
\label{sec:testing}}{Testing using randomly picking reference genes }}\label{testing-using-randomly-picking-reference-genes}}

When a subset \(S\) of independent reference genes is known, estimates
of the unkown \(s_i\), \(i=1,...,n\) are \begin{equation}
\hat s_i^S := \frac{n\sum_{j\in S}X_{ij}}{\sum_{i=1}^n\sum_{j\in S} X_{ij}}. \label{hat_si}
\end{equation} These estimates appear as a ratio of moment estimates.
Under the Poisson assumption, it is also the maximum likehood estimators
since \(\log \lambda_{ij}=\log s_i + \log \mu_j\) can be written as a
linear predictor of indicator variables for the individual \(i\) and the
gene \(j\): \[\log \lambda_{ij}=\phi_i\times I_i+\theta_j\times I_j.\]
\medskip

In practice such reference genes are unknown and it is conceptually
difficult to believe that they could be known when one new hypothesis is
tested. This problem is all the more difficult as the \(s_i\) are
unknown. Our methodology bypasses this lack of knowledge through the
sampling of the subsets \(S_1\),\ldots,\(S_r\) which are used each in
turn as normalization subset as if they were made of reference genes
(see Section \ref{sec:notations}).

For a given normalization subset \(S\) in
\(\mathcal S=\{S_1, ...,S_r\}\), our test statistic, denoted
\(T_S(X_{1 j},...,X_{n j})\), is asymptotically equivalent to
\(T^*(X_{1 j},...,X_{n j})\) with the \(s_i\) replaced by
\(\hat s_i^S\). The rejection region is adapted accordingly through a
non asymptotic control of the level by replacing \(q^*(1-\eta/2)\) with
\begin{equation}\label{eq:q}
q(1-\eta/2):=(1+\sqrt{c\log n})\,q^*(1-\eta/2)
\end{equation} where \(c\) is a positive constant. As a consequence, for
any gene with index \(j\) not in \(S\), we can determine if the null
\(H_0^j\) is rejected or not with a prescribed significance level.
\medskip

\hypertarget{our-procedure}{%
\subsection{\texorpdfstring{Our procedure
\label{sec:our-proc}}{Our procedure }}\label{our-procedure}}

For any gene \(j=1,...,m\) and any normalization subset
\(S\in \mathcal S_j\), let us define the detection indicator (the
indicator of the rejection of the null hypothesis) for gene \(j\) when
the subset \(S\) is used for normalization by
\begin{equation} \label{eq:practical-test}
\mathbbm 1_S(j):=\left(|T_S(X_{1 j},...,X_{n j})|>q(1-\eta/2)\right).
\end{equation}

We denote by \(P_j(\cdot)=P(\cdot|S\in\mathcal S_j)\) the conditionnal
probability with respect to the event \(S\in\mathcal S_j\).

Using total probability formula, we decompose the detection rate
\begin{equation}
P_j(\mathbbm 1_S(j)=1) = P_j(\mathbbm 1_S(j)=1|S\cap\mathcal D^*=\emptyset)\,P_j(S\cap\mathcal D^*=\emptyset)+P_j(\mathbbm 1_S(j)=1|S\cap\mathcal D^*\not=\emptyset)\,P_j(S\cap\mathcal D^*\not=\emptyset).
\end{equation}

\begin{table}
\centering
\begin{tabular}{c|c|c|}
\cline{2-3}
& $H_0^j$: $j$ invariant & $H_1^j$: $j$ is DE $\eta$-detectable\\
& $\mu_j^A=\mu_j^B$ & $j\in\mathcal D^*$\\\hline
\multicolumn{1}{|c|}{Good normalization subset} & \multirow{2}{*}{$\leq\eta$} & \multirow{2}{*}{$\geq 1-\beta$}\\
\multicolumn{1}{|c|}{$S\cap\mathcal D=\emptyset$} & &\\\hline
\end{tabular}
\caption{Assumptions on the detection rates for a gene $j$ regarding its status when a good normalization subset is used.\label{tab:ineq for detection rates}}
\end{table}

Assuming that the definition of \(q(1-\eta/2)\) given by (\ref{eq:q})
ensures that the detection rates for a good normalization subset satisfy
Table \ref{tab:ineq for detection rates}, it follows that \begin{align} 
P_j(\mathbbm 1_S(j)=1) &\leq \eta (1-\pi^0_{d^*}) + \pi^0_{d^*} \quad\text{ when } j\not\in\mathcal D^* \\
P_j(\mathbbm 1_S(j)=1) &\geq  (1-\beta)(1-\pi^1_{d^*}) \quad\text{ when } j\in\mathcal D^*.
\end{align} where \(\pi^0_d\) and \(\pi^1_d\) are defined in
(\ref{eq:pid}).\bigskip

Thanks to the use of a pivotal statistic,
\(P_j(\mathbbm 1_S(j)=1|S\cap\mathcal D^*=\emptyset)\) does not depend
on \(j\) hence \(\eta\) can be chosen small even when the number of
genes is large.

We consider \(R_j\) the number of detections for the gene \(j\) through
its \(r_j:=|\mathcal S_j|\) associated normalizations. Let us define
\begin{equation} \label{eq:p-values}
p_j^d(\eta) := 1-B\left(R_j;r,\kappa(\eta (1-\pi_d^0) + \pi_d^0)\right)
\end{equation} where \(\kappa=1-k/m\) (see (\ref{eq:kappa})) and
\(B(.,n,x)\) is the c.d.f. of the binomial with parameter \(n\) and
\(x\). As a consequence \(p_j^d(\eta)\) appears as the \(p\)-value
associated with \(R_j\) when \(\mathcal D^*\) is of cardinality \(d\)
and \(R_j\) is considered to come from a binomial with parameters \(r\)
and \(\kappa(\eta (1-\pi_d^0) + \pi_d^0)\).

Noticing that \(r\) and \(\kappa(\eta (1-\pi_d^0) + \pi_d^0)\) does not
depend on \(j\) the order of the \(p\)-values does not depend on
\(\eta\) or \(d\). As a consequence, we can order the genes accordingly
to the \(p_j^d(\eta)\) independently from \(d\) and \(\eta\) to satisfy
\begin{equation}p_{(1)}^d(\eta) \leq p_{(2)}^d(\eta) \leq ... \leq p_{(m)}^d(\eta), \text{ for all $d$ and $\eta$}.\end{equation}

We now derive our detection procedure:

\begin{itemize}
\item Gene $(1)$ is declared to be DE if and ony if $p_{(1)}^0(\eta)<\alpha/m$;
\item Gene $(2)$ is declared to be DE if and only if $p_{(2)}^1(\eta)<\alpha/(m-1)$;
\item ...
\item Gene $(d)$ is declared to be DE if and only if $p_{(d)}^{d-1}(\eta)<\alpha/(m-d+1)$;
\item Gene $(d+1)$ is declared to be DE if and only if $p_{(d)}^d(\eta)<\alpha/(m-d)$;
\item ...
\end{itemize}

Finally, given \(\eta>0\), we define \(\hat d\), the number of genes
declared DE by our procedure, as follows: \begin{equation}
\hat d := \left\{
\begin{array}{ccl} 
0 & \qquad & \text{ if }\quad p_{(1)}^0(\eta)\geq\alpha/m, \\
\min\left\{d>0,\, p_{(d)}^{d-1}(\eta)<\alpha/(m-d+1) \quad\text{and}\quad p_{(d+1)}^{d}(\eta)\geq\alpha/(m-d)\right\} &\qquad& \text{ otherwise.}
\end{array}
\right.
\end{equation} If \(\hat d>0\), the genes associated with the \(\hat d\)
smallest \(p\)-values are declared DE, otherwise all genes are declared
invariant. We recall that the order of the \(p\)-values does not depend
on \(d\) and \(\eta\), hence the meaning of ``smallest \(p\)-values'' is
well defined.

Clearly, our procedure will be all the more powerful that the detection
rate difference \((1-\beta)(1-\pi_d^1) - (\eta(1-\pi_d^0)+\pi_d^0)\) is
large when \(d\leq d^*\). Our main result controls the FWER of non DE
gene detection for this randomize procedure.\medskip

\begin{theo}\label{th:random-picking}
Assuming that the genes are independent and that the detection rates for the test defined by (\ref{eq:practical-test}) satisfy the assumptions provided by Table \ref{tab:ineq for detection rates} for any good normalization subset $S$, then the FWER is bounded by $\alpha + \it{o}_r(1)$ as soon as
\begin{equation}
(1-\beta)(1-\pi^1_{d^*}) > \eta (1-\pi^0_{d^*}) + \pi^0_{d^*}. \label{eq:eta-beta-ineq}
\end{equation} 
\end{theo}\medskip

The proof of Theorem \ref{th:random-picking} is reported in Appendix
\ref{fwer-control} \medskip

\hypertarget{rates-of-detection-and-power}{%
\subsection{\texorpdfstring{Rates of Detection and Power
\label{sec:power}}{Rates of Detection and Power }}\label{rates-of-detection-and-power}}

We now focus on the link between \(\eta\) and \(\alpha\). We first focus
on the case \(d^*=0\) to derive condition on \(r\) based on a level
control. Then, we study the power when \(d^*>0\). \medskip

If \(d^*=0\), the \(p_j^0(\eta)\) derived from the \(R_j\) are
associated with binomial \(\mathcal B(r,\theta_0)\) with expectation
\(r\theta_0\) where \(\theta_0 := \kappa\eta\). In this case, each
normalisation subsets \(S\) is a good normalisation subset and the rate
of detection satisfies
\(\theta_S(j):=P_j(\mathbbm 1_S(j)=1)\leq\theta_0\) such that
\(P\left(R_j>x\right) \leq P\left(\mathring R_j>x\right)\) where
\(\mathring R_j\) is a \(\mathcal B(r,\theta_0)\). This result is a
consequence of the following lemma which provides a more general result,
true for any \(d^*\).

\begin{lem} \label{lem:binomials} Denoting
$$\theta_0:=\kappa(\eta (1-\pi^0_{d}) + \pi^0_{d}) \quad \text{ and }\quad \theta_1:=\kappa(1-\beta)(1-\pi^1_{d}),$$
under assumption provided by Table \ref{tab:ineq for detection rates}, if $j_0\not \in\mathcal D^*$ then $R_{j_0}\leq \mathring R_{j_0}$ where $\mathring R_{j_0}\sim \mathcal B(r,\theta_0)$. Similarly, if $j_1 \in\mathcal D^*$ then $R_{j_1}\geq \hat R_{j_1}$ where $\hat R_{j_1}\sim \mathcal B(r,\theta_1)$. In term of c.d.f it follows that 
$$
P(R_{j_0}\leq x) \geq P(\mathring R_{j_0}\leq x) \quad \text{ and } \quad P(R_{j_1}\leq x) \leq P(\hat R_{j_1}\leq x).
$$
\end{lem}
\paragraph*{Proof of Lemma \ref{lem:binomials}}

For any \(S\) in \(\mathcal S_{j_0}\), as \(\mathbbm 1_S(j_0)\) is a
Bernoulli with parameter
\begin{equation}\theta_S(j_0):=P(\mathbbm 1_S(j_0)=1)\leq \kappa(\eta (1-\pi^0_{d}) + \pi^0_{d}) = \theta_0,\end{equation}
it follows that
\(\mathbbm 1_S(j_0) = \mathbbm 1_{U_S\leq \theta_S(j_0)} \leq \mathbbm 1_{U_S\leq \theta_0}\)
where \(U_S\) are independent uniform random variables. Consequently
\(R_{j_0} \leq \mathring R_{j_0}\) where \(\mathring R_{j_0}\) is a
Binomial \(\mathcal B(r, \theta_0)\). Similarly for
\(j_1\in\mathcal D^*\), for any \(S\) in \(\mathcal S_{j_1}\),
\(\mathbbm 1_S(j_1) = \mathbbm 1_{U_S\leq \theta_S(j_1)} \geq \mathbbm 1_{U_S\leq \theta_1}\)
with
\begin{equation}\theta_S(j_1):=P(\mathbbm 1_S(j_1)=1) \geq \kappa (1-\beta) (1-\pi^1_{d}) = \theta_1\end{equation}
and \(R_{j_1} \geq \hat R_{j_1}\) where \(\hat R_{j_1}\) is a Binomial
\(\mathcal B(r, \theta_1)\). \hfill\(\square\)\bigskip

As \(d^*=0\), the FWER is the probability under the global null,
\(\bigcap_j H_0^j\), to have one \(R_j\) too large and can be upper
bounded as follow
\begin{equation}\label{eq:link-with-tilde} P\left(\exists j\in\{1,...,m\}, R_j>r\theta_0 + r\varepsilon\right) \leq P\left(\exists j\in\{1,...,m\}, \mathring R_j>r\theta_0 + r\varepsilon\right) \leq m P\left(\mathring R_1>r\theta_0 + r\varepsilon\right). \end{equation}

Using (Massart \protect\hyperlink{ref-massart_tight_1990}{1990}, Theorem
2), \begin{equation} \label{eq:first-kind-error}
P(\mathring R_1-r\theta_0>r\varepsilon ) \leq \exp\left(-\frac{r\varepsilon^2}{2(\theta_0+\varepsilon/3)(1-\theta_0-\varepsilon/3)}\right) \leq \exp\left(-\frac{r\varepsilon^2}{2(\theta_0+\varepsilon/3)}\right).
\end{equation} This inequality is non trivial for deviations
\(r\varepsilon\) of order the standard deviation of \(\mathring R_1\)
which is of not smaller than \(\sqrt{r\theta_0}\). To go further, we
consider that the deviations statisfies
\(r\varepsilon = \sqrt{\theta_0}\,r^{0.5+\xi}\) with \(0<\xi<0.5\) such
that \begin{equation}\label{eq:eps-order}
\varepsilon = \sqrt{\theta_0}\, r^{\xi-0.5}.
\end{equation} The latter is smaller than \(\theta_0\) as soon as
\(r\geq \theta_0^{1/(2\xi-1)}\). In this case
(\ref{eq:first-kind-error}) becomes
\begin{equation} \label{eq:first-kind-error-maj}
P(\mathring R_1-r\theta_0>r\varepsilon ) \leq \exp\left(-\frac{3r\varepsilon^2}{8\theta_0}\right) \leq \exp\left(-\frac{3}{8}r^{2\xi}\right).
\end{equation}

Using (\ref{eq:link-with-tilde}) \begin{equation}
P\left(\exists j\in\{1,...,m\}, R_j>r\theta_0 + r\varepsilon\right) \leq m \exp\left({-\frac{3}{8}r^{2\xi}}\right).
\end{equation} Equating the right-hand side term with \(\alpha\), it
follows that the FWER is smaller than \(\alpha\) as soon as\\
\begin{equation}\label{eq:fwer-alpha}
r \geq \theta_0^{1/(2\xi-1)} \vee \left[-\frac 8 3 \log(\frac{\alpha}{m})\right]^{1/2\xi}.
\end{equation}\medskip

We now suppose \(d^*>0\) and assume the rate of rejections to be not
smaller than \(\theta_1\) with \(\theta_1\geq\theta_0\) for the genes
with index in \(\mathcal D^*\). Given \(j\in\mathcal D^*\), the error of
second kind can be controlled as follow:
\[P\left(R_j<r\theta_0 + r\varepsilon\right) \leq P\left(\hat R_j<r\theta_0 + r\varepsilon\right) = P\left(\hat R_j - r\theta_1< -r(\theta_1-\theta_0 - \varepsilon)\right)\]
where \(\hat R_j \sim \mathcal B(r,\theta_1)\), see Lemma
\ref{lem:binomials}. Again using Massart's inequality, it follows that
\begin{align}
P\left(\hat R_j<r\theta_0 + r\varepsilon\right) &\leq \exp \left(-\frac {r(\theta_1-\theta_0 - \varepsilon)^2} {2(1-\theta_1+(\theta_1-\theta_0 - \varepsilon)/3)(\theta_1 -(\theta_1-\theta_0 - \varepsilon)/3)} \right) \\
&\leq \exp \left(-\frac {r(\theta_1-\theta_0 - \varepsilon)^2} {2(\theta_1 -(\theta_1-\theta_0 - \varepsilon)/3)} \right)\\
&\leq \exp \left(-\frac {r(\theta_1-\theta_0 - \varepsilon)^2} {2\theta_1} \right).
\end{align} The last inequality comes from assuming
\(\varepsilon \leq \theta_1-\theta_0\). Assuming the stronger constraint
\(\varepsilon \leq (\theta_1-\theta_0)/2\), it follows than
\begin{align}
P\left(\hat R_j<r\theta_0 + r\varepsilon\right) &\leq \exp \left(-\frac {r(\theta_1-\theta_0)^2} {8\theta_1} \right) \leq \exp \left(-\frac {r(\theta_1-\theta_0)^2} {8} \right).
\end{align} Taking care of the multiplicity, it follows that
\begin{align}
P\left(\exists j\in\mathcal D^*, \hat R_j<r\theta_0 + r\varepsilon\right) &\leq d^* \exp \left(-\frac {r(\theta_1-\theta_0)^2} {8} \right)\leq m \exp \left(-\frac {r(\theta_1-\theta_0)^2} {8} \right).
\end{align}

Using (\ref{eq:eps-order}), the constraint
\(\varepsilon \leq (\theta_1-\theta_0)/2\) becomes
\[r \geq \left[\frac{2\sqrt{\theta_0}}{\theta_1-\theta_0}\right]^{1/(0.5-\xi)}.\]

Finally (using \(\xi=0.25\)) we obtain the following theorem and its
corollary which control level and power of our randomized strategy:

\begin{theo}\label{th:fwer-power}
Under the assumptions of Th. \ref{th:random-picking}, our procedure is of FWER $\alpha$ and has its power growing exponentially fast with $r$ --all the faster as $\theta_1-\theta_0$ the larger-- as soon as 
$$r \geq \frac 1 {\sqrt{\theta_0}} \vee \left[-\frac 8 3 \log(\frac{\alpha}{m})\right]^2 \vee \left[\frac{2\sqrt{\theta_0}}{\theta_1-\theta_0}\right]^4$$ with $$\theta_1-\theta_0=(1-\frac k m)\left[(1-\beta)(1-\pi^1_{d^;*}) - (\eta (1-\pi^0_{d^*}) + \pi^0_{d^*})\right].$$
\end{theo}

In other words,

\begin{cor}
Under the assumptions of Th. \ref{th:random-picking}, when $r\geq 1/\sqrt{\theta_0}$ and 
\begin{equation} \label{eq:normalized rate difference}
\frac{(1-\beta)(1-\pi^1_{d^*}) - (\eta (1-\pi^0_{d^*}) + \pi^0_{d^*})}{\sqrt{\eta (1-\pi^0_{d^*}) + \pi^0_{d^*}}} \geq 2 \frac{m}{m-k}\left(-\frac 8 3 \log \frac{\alpha}m\right)^{0.5}
\end{equation}
our procedure is of FWER $\alpha$ with power growing exponentially fast as soon as $r \geq \left[-\frac 8 3 \log(\alpha/m)\right]^{-0.5}$.
\end{cor}

The left-hand side in (\ref{eq:normalized rate difference}) appears as
lower bound of the normalized difference of detection rates between DE
and not-DE genes.

\hypertarget{testing-procedure-for-poisson-and-negative-binomial-models}{%
\section{\texorpdfstring{Testing procedure for Poisson and Negative
binomial models
\label{sec:testing-proc}}{Testing procedure for Poisson and Negative binomial models }}\label{testing-procedure-for-poisson-and-negative-binomial-models}}

Usually, the gene expressions are modelled by negative binomials
\(X_{ij}\sim \mathcal{NB}(\gamma_j^\bullet,\gamma_j^\bullet\big/(\gamma_j^\bullet+\lambda_{ij}))\)
such that \(\mathbb{E}(X_{ij})=\lambda_{ij}=s_i\mu_j^\bullet\) and
\(\textrm{var}(X_{ij})=\lambda_{ij}(1+\lambda_{ij}/\gamma_j^\bullet)\).
In this parametrization, the gene dependent parameter
\(\gamma_j^\bullet\) governs the overdispersion of the count data
\(X_{ij}\) with respect to the Poisson case for which
\(\gamma_j^\bullet=+\infty\). We will denote further
\(\rho_j^\bullet:=\mu_j^\bullet/\gamma_j^\bullet\) which is zero for the
Poisson case.

Under the assumption that \(\lambda_{ij}=s_i\,\mu_j^\bullet\) is large
enough, the Gaussian approximation of the negative binomial counts
\(X_{ij}\) provides \begin{equation}
X_{ij}\sim\mathcal {NB} (\gamma_j^\bullet,\gamma_j^\bullet\big/(\gamma_j^\bullet+\lambda_{ij}))\approx\mathcal{N}(s_i\,\mu_j^\bullet,s_i\,\mu_j^\bullet(1+s_i\,\rho_j^\bullet)) \label{eq:gauss-approxNB}
\end{equation} such that \[
2\sqrt{X_{ij}}\approx\mathcal{N}(2\sqrt{s_i\,\mu_j^\bullet},1+s_i\,\rho_j^\bullet).
\] This Gaussian approximation is just an extension of the Poisson
(\(\rho_j^\bullet\)) approximation obtained by variance stabilization.
\bigskip

Given a normalization subset \(S\), let us consider an integer
\(j\not\in S\) and let us specify the expression to be compared under
conditions \(A\) and \(B\).

Assuming that the Gaussian approximation for negative binomial holds,
from \eqref{eq:gauss-approxNB}, we obtain \begin{align}
U_{ij}:&=2\sqrt{\frac{X_{ij}}{s_i}}\approx\mathcal N(2\sqrt{\mu_j^\bullet}, \rho_j^\bullet+1/s_i) \nonumber\\
&=2\sqrt{\mu_j^\bullet}+\sqrt{\rho_j^\bullet+1/s_i}\varepsilon_{ij} = 2\sqrt{\mu_j^\bullet}+ V_{ij} \label{eq:Yij_approxNB}
\end{align} with \(\varepsilon_{ij}\sim\mathcal N(0, 1)\) and
\(V_{ij}:=\sqrt{\rho_j^\bullet+1/s_i}\varepsilon_{ij}\). By construction
the \(\varepsilon_{ij}\) for \(i=1,\ldots,n\) are independent. We denote
further \(\omega_i^\bullet := (\rho_j^\bullet +1/s_i)^{1/2}\).

Defining \(\Sigma_S^2\) as \begin{equation} \label{eq:sigma2}
\Sigma_S^2:=\sum_{i=1}^n (\omega_i^\bullet)^2=\sum_{i\in A} (\omega_i^A)^2+\sum_{i\in B} (\omega_i^B)^2=\frac{\rho_A}{n_A}+\frac{\rho_B}{n_B}+\frac 1{n_A^2}\sum_{i\in A} \frac 1{s_i}+\frac 1{n_B^2}\sum_{i\in B}\frac 1 {s_i},
\end{equation} we consider the hypotheses test statistic \[
\frac{\bar U_l^A-\bar U_l^B}{\Sigma_s} = 2\frac{\sqrt{\mu_{j}^A}-\sqrt{\mu_{j}^B}}{\Sigma_s} + \frac{\bar V_l^A-\bar V_l^B}{\Sigma_s} 
\] which, as \eqref{eq:Yij_approxNB} holds, under the null hypothesis is
reduced to \((\bar V_l^A-\bar V_l^B)/\Sigma_s\) and follows
approximately \(\mathcal{N}(0,1)\) .

Since \(\rho_j^A\), \(\rho_j^B\) and the \(s_i\) are unknown, the latter
cannot be used directly as a test statistic. Therefore we consider
further \begin{equation}
Y_{ij}:=2\sqrt{X_{ij}\big/\hat s_i} = \sqrt{s_i\big/\hat s_i} U_{il}
\end{equation} where \(\hat s_i\) is an estimator of \(s_i\), see
\eqref{hat_si}. Together with an estimator of \(\Sigma_S^2\):
\begin{equation}
\hat \Sigma_S^2 = \frac{1}{n_A(n_A-1)}\sum_{i\in A}(Y_{ij}-\bar Y_j^A)^2 + \frac{1}{n_B(n_B-1)}\sum_{i\in B}(Y_{ij}-\bar Y_j^B)^2
\end{equation} we build our testing procedure on the following statistic
\[
T:=\frac{\bar Y^A_j-\bar Y^B_j}{\hat\Sigma_S}.
\]

The justification for this construction is provided by a decomposition
of \(T\) that we obtain by using algebraic computations. First, for a
real vector \(x_j:=(x_{1j},\ldots,x_{nj})^T\) from \(\mathbb R^n\), we
denote the empirical means of \(x_j\) in \(A\) and \(B\) by \[
\bar x_l^A:=\frac 1 {n_A}\sum_{i\in A} x_{ij} \quad \text{ and } \quad \bar x_l^B:=\frac 1 {n_B}\sum_{i\in B} x_{ij}.
\]

Then we establish the relation between the difference of means and the
vector \(x_j\) \[(\bar x_j^A - \bar x_j^B) \mathbbm 1_{n}=Hx_j\] where
\[\mathbbm 1_{n} := (\underbrace{1,\ldots,1}_{n \text{ times}})^T\quad\text{ and }\quad H:=\left(\begin{array}{ll}\frac{1}{n_A}J_{n_A} & -\frac{1}{n_B}J_{n_A,n_B}\\  \frac{1}{n_A}J_{n_B,n_A} & -\frac{1}{n_B}J_{n_B}\end{array} \right).
\]

and \(J\) being the unit matrix consisting of integers equal to 1.

With respect to these notations the following decomposition of \(T\)
holds \begin{align}
T\mathbbm 1_n &= \frac{1}{\hat\Sigma_S}HY_j = \frac{\Sigma_S}{\hat\Sigma_S}\left(\frac{1}{\Sigma_S}H(Y_j-U_j)+\frac{1}{\Sigma_S}HU_j\right)\nonumber\\
&= \frac{\Sigma_S}{\hat\Sigma_S}\left(\frac{1}{\Sigma_S}H(\textrm{diag}(\sqrt{s_i/\hat s_i})U_j-U_j)+\frac{1}{\Sigma_S}H(2\sqrt{\mu_j^\bullet}+V_j) \right)\nonumber\\
&= \frac{\Sigma_S}{\hat\Sigma_S}\left(\frac{1}{\Sigma_S}H\textrm{diag}(\sqrt{s_i/\hat s_i}-1)U_j+\frac{2}{\Sigma_S}(\sqrt{\mu_j^A}-\sqrt{\mu_j^B})+\frac{\bar V_j^A-\bar V_j^B}{\Sigma_S}) \right)\nonumber\\
&= \frac{\Sigma_S}{\hat\Sigma_S}\left(\frac{1}{\Sigma_S}\underbrace{H\textrm{diag}(\sqrt{s_i/\hat s_i}-1)\textrm{diag}(\sqrt{\rho_j^\bullet+1/s_i})}_{R_1}\varepsilon_j\right.\nonumber\\
&\qquad+\frac{2}{\Sigma_S}\underbrace{H\textrm{diag}(\sqrt{s_i/\hat s_i}-1)(\sqrt{\mu_j^A}\mathbbm 1_{n_A}^T,\sqrt{\mu_j^B}\mathbbm 1_{n_B}^T)^T}_{R_2}\nonumber\\
&\qquad\left. +\frac{2}{\Sigma_S}(\sqrt{\mu_j^A}-\sqrt{\mu_j^B})+\frac{\bar V_j^A-\bar V_j^B}{\Sigma_S}) \right) \label{eq:R1-R2}\\
&=\frac{\Sigma_S}{\hat\Sigma_S}\left(\frac{R_1\varepsilon_j}{\Sigma_S}+2\frac{R_2}{\Sigma_S}+\frac 2 {\Sigma_S}(\sqrt{\mu_j^A}-\sqrt{\mu_j^B}) +\frac{\bar V_j^A-\bar V_j^B}{\Sigma_S}\right).\label{eq:stattest}
\end{align}

We recall that by construction
\((\bar V_j^A-\bar V_j^B)/\Sigma_S\sim\mathcal N(0,1)\). Moreover the
following theorem holds which controls the distribution of our test
statistic with respect to the Gaussian under the null.

\begin{theo} \label{th:control} Assuming that $\max_i\left|\sqrt{s_i/\hat s_i}-1\right|\leq 1/2$, the following relations hold with probability larger than $1-5n^{-c}$
$$
\frac{\Sigma_S}{\hat\Sigma_S} \leq (1+\sqrt{c\log n})(1+2\max_i\left|\sqrt{s_i/\hat s_i}-1\right|)\leq (1+\sqrt{c\log n})\left(1+2\left[ 2(1+c)(1+o(1))\frac{1+s_{\max}\bar\rho_S}{\sum_{j\in S}\mu_j}\frac{\log n}n\right]^{1/2}\right),
$$
$$
\frac{\|R_1\varepsilon\|^2}{\Sigma_S^2} \leq {2(1+2\sqrt{c\log n} +2c\log n)(1+c)(1+o(1))}\frac{1+s_{\max}\bar\rho_S}{{\sum_{j\in S}\mu_j}}\log n,
$$

$$
\frac{\|R_2\|^2}{\Sigma_S^2}\leq 2(1+c)(1+o(1))\left(\sqrt{\mu_j^A}+\sqrt{\mu_j^B}\right)^2 \times\frac{1+s_{\max}\bar\rho_S}{{\sum_{j\in S}\mu_j}}(\frac n{n_A}\vee\frac{n}{n_B})\,\log n.
$$

\begin{equation}\label{eq:smax}
s_{\max}:=\max_{i=1,...,n} s_i \quad\text{ and }\quad\bar\rho_S:=\sum_{j\in S}\mu_j\rho_j\big/\sum_{j\in S}\mu_j.
\end{equation}
\end{theo}

To understand the meaning of these inequalities, one has to think that
\(\sum_{j\in S}\mu_j\) is intended to be large. As a consequence,
\(\Sigma_S/\hat\Sigma_S\) is of order \(1+\sqrt{c\log n}\) for any
\(c>0\) and the bias terms \(R_1\varepsilon/\Sigma_S\) and
\(R_2/\Sigma_S\) are negligeable. Section \ref{sec:empirical} discusses
strategies to ensure that \(\sum_{j\in S}\mu_j\) is large. \medskip

The proof of this theorem is reported in Appendix
\ref{sec:proof distrib} Let us denote \(\Phi\) the cumulative
distribution function of the standard Gaussian.

\begin{cor} \label{cor:control}
Under assumption of the Theorem \ref{th:control}, $T_S^{*}(X_{\cdot j})\approx\frac{\Sigma_S}{\hat \Sigma_S}T^{0}(X_{\cdot j})$. Moreover, if we consider $q^{*}(1-\alpha/2) = (1+\sqrt{c\log n})\Phi^{-1}(1-\alpha/2)$, then the test defined by the rejection region 
$$|T_S^{*}(X_{\cdot j})| > q^{*}(1-\alpha/2)$$ is of level $\alpha$.

\end{cor}

\bigskip

\hypertarget{implementation-strategies}{%
\section{\texorpdfstring{Implementation strategies
\label{sec:implementation}}{Implementation strategies }}\label{implementation-strategies}}

The technical assumption given by (\ref{eq:eta-beta-ineq}) constraints
\(\hat d\) to be not larger than a quantity \(\Delta\) which depends on
\(\eta\), \(\beta\), \(k\) and \(m\) and is given by
\[\Delta = \max_{d} \{d\, /\, (1-\beta)(1-\pi^1_{d}) > \eta (1-\pi^0_{d}) + \pi^0_{d} \}.\]
It expresses that the number of genes which can be detected is upper
bounded. In order to bypass this constraint on the maximal number of
possible detections, we propose an iterative implementation of our
procedure which still controls the FWER at level
\(\alpha + \it{o}_r(1)\). It is built as follow. Starting from the \(m\)
genes, the procedure is run at level \(\alpha/2\). At the \(i\)-th step,
if the number of detections reaches the upper bound \(\Delta\) (computed
at each step), then the remaining non DE-detected genes are tested using
\(\alpha/2^{i+1}\) instead of \(\alpha\). The global FWER is then
controlled by \(\sum_{i=1}^\infty \alpha/2^i\leq\alpha\).\medskip   

One essential term in the control provided by Theorem \ref{th:control}
is \(\sum_{j\in S}\mu_j\) which should be large. We can imagine two
implementations to ensure this property. The first one, which is the
closest to our theoretical setting, consists in fixing a minimal
expected intensity \(\mu_0\) per gene and select the subset made of
those genes satisfying \(\mu_j\geq\mu_0\). Along the iterations, the
normalizing subset \(S\) is such that
\(\sum_{j\in S}\mu_j\geq |S|\times\mu_0\). The second consists, given a
fix \(M_0\), in growing at each iteration \(S\) until
\(\sum_{j\in S}\mu_j\geq M_0\).

Both strategies depend on the knowledge of the \(s_i\) however, using
the estimates \(\hat s_i\) obtained from \(S\), it is possible to get an
idea whether the size of \(S\) is large enough or not by using, for
example,
\[\left(\frac 1 n\sum_{i=1}^n \sqrt{Y_{ij}\big / \hat s_i} \right)^2 \approx \mu_j \qquad \text{ and }\qquad \sum_{j\in S}\left(\frac 1 n\sum_{i=1}^n \sqrt{Y_{ij}\big / \hat s_i} \right)^2 \approx \sum_{j\in S} \mu_j.\]
In the second strategy, the growth of \(S\) stops as soon as the latter
left hand side term is larger than \(M_0\).

\bigskip

\hypertarget{empirical-study}{%
\section{\texorpdfstring{Empirical study
\label{sec:empirical}}{Empirical study }}\label{empirical-study}}

For the empirical study, we use our iterative procedure as described in
Section \ref{sec:implementation}. We consider two populations having
equal size \(n/2\) assuming that all experiments perform perfectly
(\(s_i=1\) for \(i=1,\ldots,n\)) and that all genes follow a Poisson
distribution (\(\rho_A=\rho_B=0\)).

In population \(A\), \(\mu_j^A=\mu_0\) for all \(j\). In population
\(B\), \(\mu_j^B=\mu_j^A=\mu_0\) for \(j>m_1\) and
\(\mu_j^B=\mu_0\varphi\) for \(j\leq m_1\).

According to \eqref{eq:R1-R2}, for the differentially expressed genes
(\(j\leq m_1\)), up to the biais terms \(R_1\) and \(R_2\) and up to the
term \(\Sigma_S/\hat \Sigma_S\), the expected fold change is
\(2(\sqrt{\mu_j^B}-\sqrt{\mu_j^A})/ \Sigma_S\). As
\(\Sigma_S=2/\sqrt n\), the test statistic \(T\) is of order
\(\sqrt {n\mu_0} \, |\sqrt{\varphi}-1|\) and it should be compared to
\(-q_{\alpha/2m}(1+\sqrt{c\log n})\), where \(q_{\alpha/2m}\) is the
quantile of order \(\alpha/2m\) of the standard Gaussian.

From the relation
\(\sqrt {n\mu_0} \, |\sqrt{\varphi}-1|>-q_{\alpha/2m}(1+\sqrt{c\log n})\),
we compute the lower and upper threshold values of \(\varphi\) to have
detections with probability larger than \(\alpha\) : \[ 
\begin{aligned}
\varphi_{low} &=\big(1-\frac{q_{\alpha/2m}(1+\sqrt{c\log n})}{\sqrt {n\mu_0}}\big)^2 \\
\varphi_{up} &=\big(1+\frac{q_{\alpha/2m}(1+\sqrt{c\log n})}{\sqrt {n\mu_0}}\big)^2. 
\end{aligned}
\]

In all examples, we fix \(n=\) 12, \(m=\) 500, \(\mu_0=\) 100, \(\eta=\)
5 \%, \(\alpha=\) 5 \%, \(\beta=\) 10 \% and tune \(c=\) 2 in Theorem
\ref{th:control} and its Corrolary \ref{cor:control} to ensure a proper
FWER control. For each setting, 100 data samples are simulated for two
populations \(A\) and \(B\). \medskip

We start our empirical study by considering the FWER, taking
\(\mu_j^A=\mu_j^B=\mu_0\) for all \(j\). Then we study empirically the
power of our procedure for various values of the shift \(\varphi\)
between population \(A\) and \(B\) for the genes \(j\) with
\(j=1,...,225\).

\hypertarget{under-the-global-null}{%
\subsection{Under the global null}\label{under-the-global-null}}

Here we assume that we are under the global null hypothesis for which
\(\mu_j^A=\mu_j^B=\mu_0\) for \(j=1,..., m\) and we run 2 complete sets
of simulations.\bigskip

\begin{center}\includegraphics[width=0.8\linewidth]{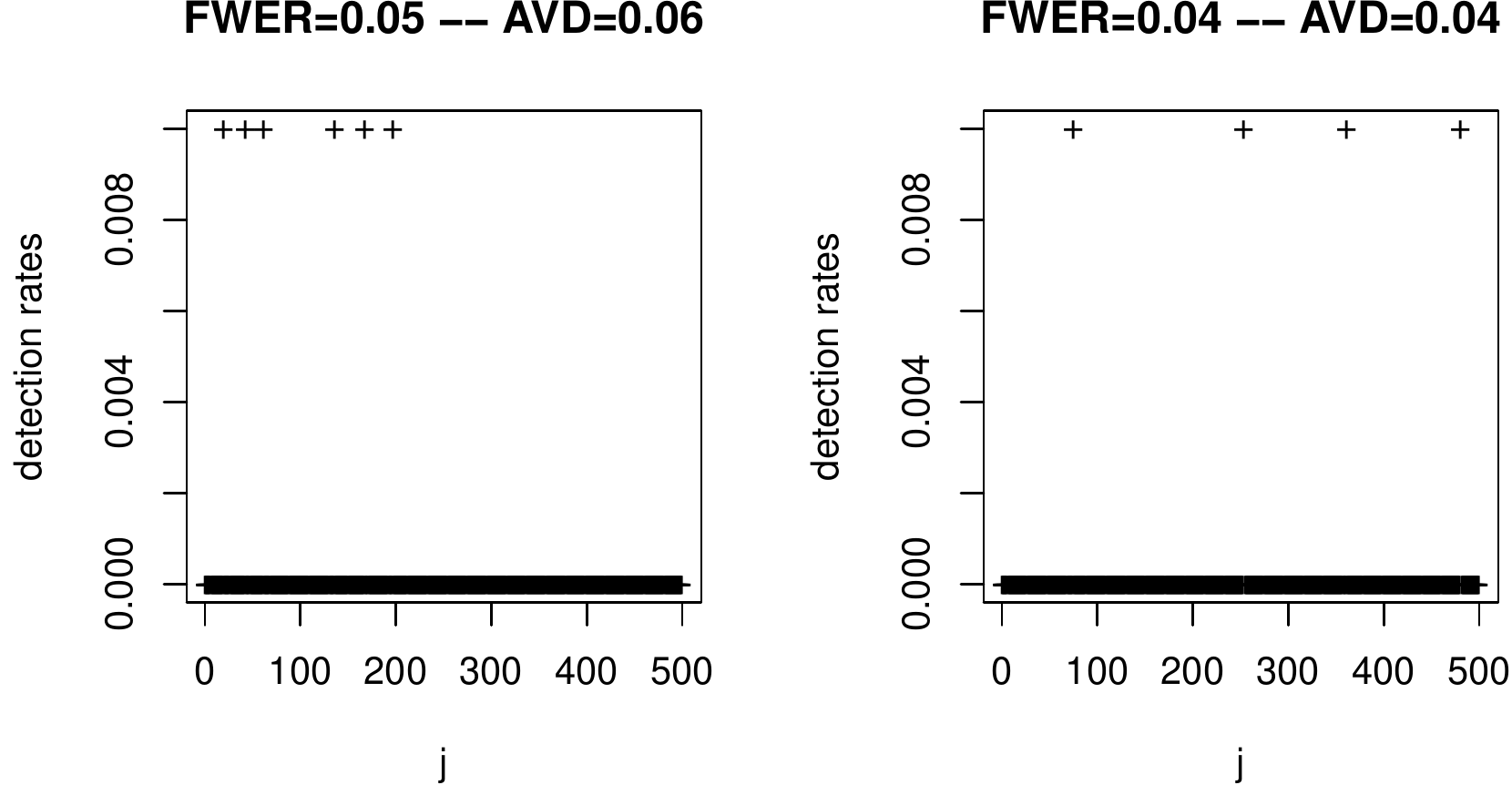} \end{center}

We observe on the first simulation (left graph) that the FWER is 0.05
with 0.06 average detected genes. For the second simulation (right
graph) FWER and average detected genes are respectively 0.04 and
0.04.\bigskip

\hypertarget{empirical-study-of-the-power}{%
\subsection{Empirical study of the
power}\label{empirical-study-of-the-power}}

We now study the power of our procedure by assuming that \(m_1=\) 225
and, while \(\mu_J^B=\mu_j^A\) for \(j>m_1\), for \(j\leq m_1\),
\(\mu_j^B=\varphi_j\mu_j^A\) with \(\phi_j := 1+a/\sqrt{j}\) such that
the relative decay \((\mu_j^B-\mu_j^A)/\mu_j^A\) varies as
\(a/\sqrt{j}\). Using \(a = 5, 7, 10\) and \(12\), we let the number of
expected detections vary from low to high and study the behavior of our
iterative procedure when the number of potential detections is
approaching \(m/2\). As \(15^2=225\), the largest relative decay is
\(a/15\). The detection rates through simulations are depicted in each
graph. On each plot below, the red dashed line specifies \(j\)
corresponding to the threshold value of \(\varphi_{up}\) when
\(\mu_1\geq\mu_0\). We observed that, as soon as the relative decay is
larger than 2/3 (\(a=10\)), we achieved an almost perfect detection.

\begin{center}\includegraphics[width=0.8\linewidth]{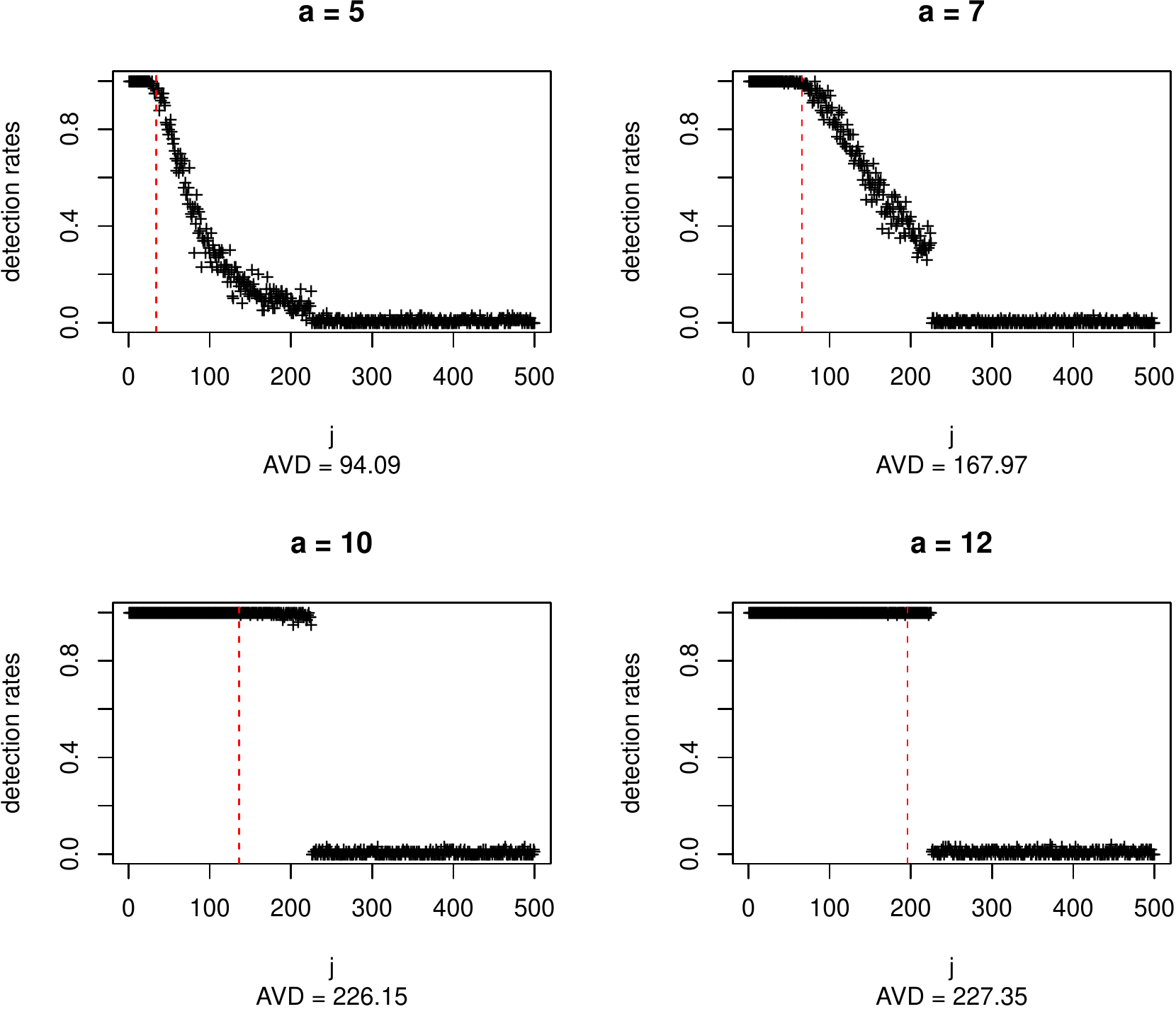} \end{center}

\hypertarget{real-data---mice-model-of-nash}{%
\section{\texorpdfstring{Real data - Mice model of NASH
\label{sec:real-data}}{Real data - Mice model of NASH }}\label{real-data---mice-model-of-nash}}

We apply the iterative procedure, as described in Section
\ref{sec:implementation}, using the same settings as in the Section
\ref{sec:empirical}, on the real data set of hepatic miRNA expressions
from a mouse model of non-alcoholic steatohepatitis (NASH), the CDAHFD
(choline-deficient, L-amino acid-defined, high-fat diet) model, with 4
cases of NASH with hepatic fibrosis and 4 controls without hepatic
lesions (Hoffmann et al.
\protect\hyperlink{ref-hoffmann_hepatic_2020}{2020}). One mouse showed a
strong disequilibrium of library size with respect to the others of
about a third.

Expressions with less than 20 reads shown in total in the 8 mice were
filtered out prior to analysis and the testing was done for 749
remaining miRNA's. For these remaining genes, if their are not
differentially expressed, then \(\mu_0 \approx\) 2.5 (see Section
\ref{sec:implementation}). For all 2500 randomizations, normalization
subsets were formed with 10 random reference genes.

The results achieved on this real experiments were compared to those
obtained using the \emph{trimmed mean of M values} (TMM) (Mark D.
Robinson and Oshlack
\protect\hyperlink{ref-robinson_scaling_2010}{2010}).

The procedure detectes the total of 14 miRNA differentially expressed:
mmu-miR-31-3p, mmu-miR-31-5p, mmu-miR-34a-5p, mmu-miR-96-5p,
mmu-miR-141-3p, mmu-miR-141-5p, mmu-miR-200a-3p, mmu-miR-200b-3p,
mmu-miR-200b-5p, mmu-miR-200c-3p, mmu-miR-429-3p, mmu-miR-582-5p,
mmu-miR-802-3p, mmu-miR-802-5p.

Within the mmu-miR-34 family members that are known to regulate hepatic
fibrosis (Li et al. \protect\hyperlink{ref-li_microrna-34a_2015}{2015};
Jiang et al. \protect\hyperlink{ref-jiang_roles_2017}{2017}), only the
major isoform mmu-miR-34a-5p whose monocistronic gene is located on
chromosome 4 has been detected and identified strongly upregulated while
the other major isoforms mmu-miR-34b-5p and mmu-miR-34c-5p whose
polycistronic gene is located on chromosome 9 are not. All other minor
microRNA isoforms mmu-miR-34a/b/c/-3p does not exhibit significant
differences in the reads between the experimental groups.

The microRNA mmu-miR-96-5p is involved in both fibrosis (Chandel et al.
\protect\hyperlink{ref-chandel_association_2018}{2018}) and
cancerization processes (Leung et al.
\protect\hyperlink{ref-leung_wnt-catenin_2015}{2015}), showing that
there are possible linkages between these different pathologies. In our
experimental model of NASH we observe a clear increase in its
expression, however it is also consistant with the fact that nodules
were observed in the livers of mice at the time of sacrifice.

For mmu-miR-34a-5p and mmu-miR-96-5p, due to the clear increase in
expression observed both our procedure and TMM behave the same. For the
further miRNA, TMM clearly shows a lower sensibility.

Our differential expression analysis also identified a subset of 7
microRNAs that had significant differences between the experimental
groups. These belong to the mmu-miR-200 family, which is known for its
involvement in various liver diseases including NASH, fibrosis, and
hepatocellular carcinoma (Murakami et al.
\protect\hyperlink{ref-murakami_progression_2011}{2011}; Gregory et al.
\protect\hyperlink{ref-gregory_mir_200_2008}{2008}; Jiang et al.
\protect\hyperlink{ref-jiang_roles_2017}{2017}). The mmu-miR-200 family
is divided into two different clusters: one of the clusters is located
on chromosome 4, which contains mmu-miR-200a/b/429 members, the other
cluster which is located on chromosome 6 includes mmu-miR-200c/141
members. Although belonging to different clusters located in different
chromosomes, the major microRNA isoforms (mmu-miR-141-3p,
mmu-miR-200a-3p, mmu-miR-200b-3p, mmu-miR-200c-3p, mmu-miR-429-3p) are
all overexpressed in the CDAHFD model, and the same evolution is
observed with the trimmed mean of M-values normalization method (TMM
normalization method). On the other hand, our procedure is able to show
that minor microRNA isoforms (mmu-miR-141-5p, mmu-miR-200b-5p) are also
upregulated in the CDAHFD model, which is not observed with TMM.

Interestingly, mmu-miR-31-5p which was previously shown overexpressed in
liver fibrosis (Hu et al. \protect\hyperlink{ref-hu_role_2015}{2015}),
cirrhosis, and hepatocellular carcinoma (Karakatsanis et al.
\protect\hyperlink{ref-karakatsanis_expression_2013}{2013}; Tessitore et
al. \protect\hyperlink{ref-tessitore_microrna_2016}{2016}) is found
upregulated in the CDAHFD model. In this case our procedure is able to
show that its minor microRNA counterpart mmu-miR-31-3p is also
upregulated. Here again this result is not observed with TMM.

Finally, the last three microRNAs (mmu-miR-582-5p (Zhang et al.
\protect\hyperlink{ref-zhang_mir-582-5p_2015}{2015}), mmu-miR-802-3p,
mmu-miR-802-5p (Zhen et al.
\protect\hyperlink{ref-zhen_microrna-802_2018}{2018}; Yang et al.
\protect\hyperlink{ref-yang_microrna-802_2019}{2019})) are intriguing
because their known activities are not specific to the field of
fibrosis, so further investigation is required to understand their
involvement in the pathology. It should be emphasized that the increase
in their expression involves a relatively small number of reads and are
not detected by TMM.

All this biological arguments show that our method is more sensitive
than those used in daily practice by biologists and offer some
meaningful results.

From a statistical point-of-view, as our procedure is based on some
random sampling, it depends on the seed of the random generator. To get
an idea of the stability with respect to this seed, we run few times the
complete procedure on our real data. On the 14 detected miRNA, only
mmu-miR-96-5p was not present in the list of detected on few runs.

\hypertarget{appendix}{%
\section{Appendix}\label{appendix}}

\hypertarget{fwer-control-of-randomly-picking-reference-genes}{%
\subsection{FWER control of randomly picking reference
genes}\label{fwer-control-of-randomly-picking-reference-genes}}

\label{fwer-control}

We recall that the \(p_j^d(\eta)\) defined in (\ref{eq:p-values})
correspond to the \(p\)-values associated to the numbers of detections
\(R_j\) when \(d^*=d\) and the error of type I is control by \(\eta\)
for a good normalization subset. \medskip

In order to establish that our randomized procedure respects FWER at
level \(\alpha\), we first yield the following lemma which controls the
order between DE genes and non-DE genes of the \(p_j^d(\eta)\).

\begin{lem} \label{lem:p-values} If the genes are independent and the following relation holds $$P_{j\not\in \mathcal D^*}(\mathbbm 1_S(j)=1)\leq\eta (1-\pi^0_{d^*}) + \pi^0_{d^*}<(1-\beta)(1-\pi^1_{d^*})\leq P_{j\in \mathcal D^*}(\mathbbm 1_S(j)=1) $$ then, for all $d$, with small probability the $p_j^d(\eta)$ for DE genes are larger than those of non-DE genes, that is:
$$P\left(\min_{j_0\not\in\mathcal D^{*}} p_{j_0}^d(\eta) \,<\, \max_{j_1\in\mathcal D^{*}} p_{j_1}^d(\eta)\right) = \it{o}_{r}(1)\,.$$
\end{lem}

Given this lemma, with large probability, the discoveries of the DE
genes are first and false discoveries occurs if the number of
discoveries is larger than \(d^*\). As a consequence, with large
probability, a false discovery occurs if and only if
\[p_{(d^*+1)}^{d^*}(\eta) < \frac{\alpha}{m-d^*}.\] In other words, with
large probability, a false discovery occurs if and only if one of the
\(m-d^*\) non-DE genes is detected at the corrected \(\alpha/(m-d^*)\)
level which is the \emph{ad-hoc} Holm's correction to control the FWER.
Finally, the FWER is controled by \(\alpha+ \it{o}_{r}(1)\).

\paragraph*{Proof of Lemma \ref{lem:p-values}:}

From Lemma \ref{lem:binomials}, for all \(j_0\not\in\mathcal D^*\) and
all \(j_1\in\mathcal D^*\), we have \(\hat R_{j_1} \leq R_{j_1}\) and
\(R_{j_0} \leq \mathring R_{j_0}\) with
\(\mathring R_{j_0}\sim\mathcal B(r,\theta_0)\) and
\(\hat R_{j_1}\sim\mathcal B(r,\theta_1)\). As a consequence, if
\(\max_{j_0\not\in\mathcal D^{*}} R_{j_0} \,>\, \min_{j_1\in\mathcal D^{*}} R_{j_1}\)
then
\(\max_{j_0\not\in\mathcal D^{*}} \mathring R_{j_0} \,>\, \min_{j_1\in\mathcal D^{*}} \hat R_{j_1}\).
As a consequence, the following relation holds \begin{equation}
P\left(\max_{j_0\not\in\mathcal D^{*}} R_{j_0} \,>\, \min_{j_1\in\mathcal D^{*}} R_{j_1}\right) \leq P\left(\max_{j_0\not\in\mathcal D^{*}} \mathring R_{j_0} \,>\, \min_{j_1\in\mathcal D^{*}} \hat R_{j_1}\right)
\end{equation} and it is enough to prove that
\(\max_{j_0\not\in\mathcal D^{*}} \mathring R_{j_0} \,>\, \min_{j_1\in\mathcal D^{*}} \hat R_{j_1}\)
occurs with small probability when \(r\) goes to infinity. Using total
probability formula, for all \(x\) \[
\begin{aligned}
P\left(\max_{j_0\not\in\mathcal D^{*}} \mathring R_{j_0} \,>\, \min_{j_1\in\mathcal D^{*}} \hat R_{j_1}\right) &= P(\max_{j_0\not\in\mathcal D^{*}}\mathring R_{j_0}> \min_{j_1\in\mathcal D^{*}}\hat R_{j_1}|\max_{j_0\not\in\mathcal D^{*}}\mathring R_{j_0}\leq x)P(\max_{j_0\not\in\mathcal D^{*}}\mathring R_{j_0}\leq x)\\ &+P(\max_{j_0\not\in\mathcal D^{*}}\mathring R_{j_0}> \min_{j_1\in\mathcal D^{*}}\hat R_{j_1}|\max_{j_0\not\in\mathcal D^{*}}\mathring R_{j_0}> x)P(\max_{j_0\not\in\mathcal D^{*}}\mathring R_{j_0}> x)\\
&\leq P(\min_{j_1\in\mathcal D^{*}}\hat R_{j_1}\leq x)+ P(\max_{j_0\not\in\mathcal D^{*}}\mathring R_{j_0}>x)\\
&= P(\min_{j_1\in\mathcal D^{*}}\hat R_{j_1}\leq x)+ 1-P(\max_{j_0\not\in\mathcal D^{*}}\mathring R_{j_0}\leq x).
\end{aligned}
\]

As \(|\mathcal D^*|=d^*\), usual derivations of the c.d.f. for the
minimum and the maximum of independent variables provide \[
\begin{aligned}
P\left(\max_{j_0\not\in\mathcal D^{*}} \mathring R_{j_0} \,>\, \min_{j_1\in\mathcal D^{*}} \hat R_{j_1}\right) &= 1-[1-P(\hat R_{j_1}\leq x)]^{d^*}+1-[P(\mathring R_{j_0}\leq x)]^{m-d^*}\\
&= 1-[1-P(\hat R_{j_1}\leq x)]^{d^*}+1-[1-P(\mathring R_{j_0}> x)]^{m-d^*}\\
&\approx 1-(1-d^*P(\hat R_{j_1}\leq x))+1-(1-(m-d^*)P(\mathring R_{j_0}> x))\\
&= d^*P(\hat R_{j_1}\leq x)+(m-d^*)P(\mathring R_{j_0}> x)
\end{aligned}
\]

For any \(0<\xi<1\), let \(x:=B^{-1}(1-\xi,r,\theta_0)\) be the
\((1-\xi)\)-quantile under \(\mathcal B(r, \theta_0)\). Then
\(P(\mathring R_{j_0}> x)\leq \xi\). If
\(P(\hat R_{j_1}\leq x) \leq \xi\) also holds, that is if
\begin{equation}\label{eq:order of quantiles}
B^{-1}(1-\xi,r,\theta_0) \leq B^{-1}(\xi,r,\theta_1),
\end{equation} then \begin{equation} \label{eq:m-xi}
P\left(\max_{j_0\not\in\mathcal D^{*}} R_{j_0} \,>\, \min_{j_1\in\mathcal D^{*}} R_{j_1}\right) \leq P\left(\max_{j_0\not\in\mathcal D^{*}} \mathring R_{j_0} \,>\, \min_{j_1\in\mathcal D^{*}} \hat R_{j_1}\right) \leq m\xi.
\end{equation}

We now prove that \(\xi\) can be chosen satisfying
\(P(\hat R_{j_1}\leq x) \leq \xi\) if \(r\) is large enough. From the
Massart'90 inequality, we obtain that
\(B^{-1}(\xi,r,\theta_1)\geq r(\theta_1-\hat\varepsilon)\) with \[
P(\hat R_{j_1} <r(\theta_1-\hat\varepsilon))\leq \exp(-\frac{r\hat \varepsilon^2}{2(1-\theta_1+\hat \varepsilon/3)(\theta_1-\hat \varepsilon/3)})\leq \exp(-\frac{r\hat \varepsilon^2}{2\theta_1}), \quad j_1\in\mathcal D^{*}
\] similarly
\(B^{-1}(1-\xi,r,\theta_0) \leq r(\theta_0+\mathring \varepsilon)\) with
\[
P(\mathring R_{j_0}>r(\theta_0+\mathring \varepsilon))\leq \exp(-\frac{r\mathring \varepsilon^2}{2(\theta_0+\mathring \varepsilon/3)(1-\theta_0-\mathring \varepsilon/3)})\leq \exp(-\frac{r\mathring \varepsilon^2}{2(\theta_0+\mathring \varepsilon/3)}), \quad j_0\not\in\mathcal D^{*}.
\] Let us assume from now on that \(\mathring \varepsilon\) and
\(\hat \varepsilon\) are not larger than \((\theta_1-\theta_0)/2\), then
\[B^{-1}(1-\xi,r,\theta_0) \leq r(\theta_0+\mathring \varepsilon) \leq r(\theta_1 - \hat \varepsilon) \leq B^{-1}(\xi,r,\theta_1).\]
Moreover \(\mathring \varepsilon/3\leq \theta_1-\theta_0\) such that \[
P(\mathring R_{j_0}>r(\theta_0+\mathring \varepsilon))\leq  \exp(-\frac{r\mathring \varepsilon^2}{2\theta_1}), \quad j_0\not\in\mathcal D^{*}.
\] Consequently
\(P(\mathring R_{j_0}> B^{-1}(1-\xi,r,\theta_0))\leq \xi\) and
\(P(\hat R_{j_1}\leq B^{-1}(\xi,r,\theta_1)) \leq \xi\) are satisfied as
soon as \(\hat \varepsilon\) and \(\mathring \varepsilon\) are non
smaller than \(\sqrt{-2\theta_1\ln(\xi)\big /r}\). The assumption that
\(\mathring \varepsilon\) and \(\hat \varepsilon\) are not larger than
\((\theta_1-\theta_0)/2\) is then fulfilled as soon as
\[r \geq \frac{-8\theta_1\ln{\xi}}{(\theta_1-\theta_0)^2}.\] Replacing
\(\xi\) by \(\xi/m\) ends the demonstration of the lemma.
\hfill \(\square\)

\hypertarget{proof-of-theorem}{%
\subsection{\texorpdfstring{Proof of Theorem
\ref{th:control}}{Proof of Theorem }}\label{proof-of-theorem}}

\label{sec:proof distrib}

In order to derive the distribution of our test statistic, we need the
following lemma and its corrolary to control the deviations of
\(\max_i|\sqrt{s_i/\hat s_i}-1|\).

\begin{lem} \label{lem:ineqNB} Let $\hat s_i$ be the estimator of $s_i$ given by \eqref{hat_si}. Assuming the Gaussian approximations hold, then if $t^2$ is of smaller order than $n\sum_{i\in S}\mu_j\big/(1+s_{\max}\,\bar\rho_S)$, with probability larger than $1-4n\exp (-t^2/2)$,
\begin{equation}
\left|\sqrt{\frac{s_i}{\hat s_i}}-1\right|  \leq (1+o(1))t\sqrt{\frac{1+s_{\max}\bar\rho_S}{n\sum_{j\in S}\mu_j}}\qquad\text{ for all } i=1,\ldots,n. 
\label{eq:hat si controlNB}
\end{equation}

\end{lem}

\begin{cor}
Taking $t^2=2(1+c)\log(n)$, assuming that $(1+s_{\max}\,\bar\rho_S)\log(n)\big/n\sum_{i\in S}\mu_j$ is of smaller order than 1, we have with probabilty larger than $1-4n^{-c}$ for all $i=1,...,n$
\begin{equation} \label{eq:cor}
\left(\sqrt{\frac{s_i}{\hat s_i}}-1\right)^2 \leq {2(1+c)(1+o(1))}\frac{1+s_{\max}\bar\rho_S}{{\sum_{j\in S}\mu_j}}{\frac{\log n}{n}}.
\end{equation}
\end{cor}

If \(\sum_{j\in S}\mu_j\geq 1+ s_{\max}\,\bar\rho_S\), \eqref{eq:cor} is
always satisfied. This condition means that the sum of the absolute
levels over the normalization set is large enough to compensate for
over-dispersion.

The proof of Lemma \ref{lem:ineqNB} in reported in Appendix
\ref{proof:lemma}

\hypertarget{control-of-sigma_sbighat-sigma_s}{%
\subsubsection{\texorpdfstring{Control of
\(\Sigma_S\big/\hat \Sigma_S\)}{Control of \textbackslash Sigma\_S\textbackslash big/\textbackslash hat \textbackslash Sigma\_S}}\label{control-of-sigma_sbighat-sigma_s}}

Let us consider the operator \(\mathbb S(\cdot)\) defined for all
\(x\in\mathbb R^n\) by \(\mathbb S^2(x) := \|L(I-K)x\|^2\) where
\[K=\left(\begin{array}{ll}\frac{1}{n_A}J_{n_A} & 0_{n_A,n_B}\\  0_{n_B,n_A} & \frac{1}{n_B}J_{n_B}\end{array} \right)\]
and \(L\) is the diagonal matrix
\(\textrm{diag}(1/\sqrt{n_A(n_A-1)},\ldots,1/\sqrt{n_A(n_A-1)},1/\sqrt{n_B(n_B-1)},\ldots,1/\sqrt{n_B(n_B-1)})\).
Simple computations show that \(\hat \Sigma_S^2 = \mathbb S^2(Y)\).

The following relation holds \[
\begin{aligned}
\mathbb S^2(U) &=  \|L(I-K)U\|^2 = \|L(I-K)\textrm{diag}(1-\sqrt{s_i/\hat s_i})U + L(I-K)\textrm{diag}(\sqrt{s_i/\hat s_i})U\|^2
\end{aligned}
\] and using triangular inequality, as
\(Y_{ij}=\sqrt{s_i/\hat s_i}U_{ij}\), we obtain
\begin{equation}\label{eq:SU-SY}
\mathbb S(U) \leq \|L(I-K)\textrm{diag}(1-\sqrt{s_i/\hat s_i})U\| + \|L(I-K)Y\| \leq \max_i\left|\sqrt{s_i/\hat s_i}-1\right|\,\mathbb S(U) + \mathbb S(Y).
\end{equation}

It is clear that operator \(\mathbb S(\cdot)\) is invariant by any
translation of a vector which is constant over the indexes in \(A\) and
in \(B\), which is the case for \(E(U)\). As a consequence,
\(\mathbb S(U) = \|L(I-K)\textrm{diag}(\omega_i)\varepsilon\|\) where
\(\omega_i^2=\rho_j^\bullet+1/s_i\) satifies
\(U_{il}=2\sqrt{\mu_j^\bullet}+\omega_i\varepsilon_{ij}\). We control
this latter norm using Gendre
(\protect\hyperlink{ref-gendre_model_2014}{2014} Lemma 8.2).

Let us denote \(A:=L(I-K)\textrm{diag}(\omega_i)\), we now show that
\(\textrm{Tr}(AA^T)=\Sigma_S^2\) and bound the ratio
\(\Sigma_S^2\big/\mathbb S^2(Y)\) from above: \[
AA^T =  L(I-K)\textrm{diag}(\omega_i)\textrm{diag}(\omega_i)^T (I-K)^T L^T = L(I-K)\textrm{diag}(\omega_i^2)(I-K)L.
\] Since \(I-K\) and \(L\) are both symmetric, all the matrices in this
latter product are symmetric such that we can use any permutation of
them to compute the trace of this product. Hence
\(\textrm{Tr}(A A^T)=\textrm{Tr}(L^2(I-K)^2\textrm{diag}(\omega_i^2))\).

By definition
\(L^2=\textrm{diag}(1\big/n_A(n_A-1),\ldots,1\big/n_A(n_A-1),1\big/n_B(n_B-1),\ldots, 1\big/n_B(n_B-1))\)
and, as \(K\) is a matrix of projection \(K^2=K\), the same holds for
\(I-H\) such that \((I-K)^2=I-K\).

We write \(L^2(I-K)\textrm{diag}(\omega_i^2)\) as a block matrix
\[L^2(I-K)\textrm{diag}(\omega_i^2)=\left(\begin{array}{ll}\frac{1}{n_A(n_A-1)}(I_{n_A}-\frac{1}{n_A}J_{n_A})\textrm{diag}(\frac{1}{ s_i}+\rho_{j}^A) & 0_{n_A,n_B}\\  0_{n_B,n_A} & \frac{1}{n_A(n_A-1)}(I_{n_B}-\frac{1}{n_B}J_{n_B})\textrm{diag}(\frac{1}{ s_i}+\rho_{j}^B)\end{array} \right)\]
whose trace is
\(\frac{1}{n_A^2}\sum_{i\in A}(\frac{1}{s_i}+\rho_{j}^A)+\frac{1}{n_B^2}\sum_{i\in B}(\frac{1}{s_i}+\rho_{j}^B)\)
which is \(\Sigma_S^2\).

Clearly, for all \(x\in\mathbb R^n\), \(x^T A^T A x=\|Ax\|^2\geq 0\),
hence the matrix \(A A^T\) is positive and has all its eigenvalue non
negative which are all smaller than \(\textrm{Tr}(AA^T)=\Sigma_S^2\). It
follows from Gendre (\protect\hyperlink{ref-gendre_model_2014}{2014}
Lemma 8.2) that, with probability larger than \(1-e^{-x}\)
\[1-2\sqrt x\leq \frac{\mathbb S^2(U)}{\Sigma_S^2}\] or in other terms
\begin{equation} \label{eq:sigma-SU}
\frac{\Sigma_S}{\mathbb S(U)}\leq \frac{1}{\sqrt{1-2\sqrt x}}\approx 1+\sqrt x.
\end{equation}

Multiplying \eqref{eq:SU-SY} and \eqref{eq:sigma-SU} together, we get \[
\frac{\Sigma_S}{\mathbb S(Y)} \leq \frac{1+\sqrt x}{1-\max_i\left|\sqrt{s_i/\hat s_i}-1\right|}.
\] Assuming moreover that the maximum is smaller than 1/2, we can use
that \(1/(1-x)\leq 1+2x\) for \(x\leq 1/2\), such that Lemma 1 provides
\[
\frac{\Sigma_S}{\mathbb S(Y)} \leq (1+\sqrt x)(1+2\max_i\left|\sqrt{s_i/\hat s_i}-1\right|)\leq (1+\sqrt x)\left(1+2\left[ 2(1+c)(1+o(1))\frac{1+s_{\max}\bar\rho_S}{\sum_{j\in S}\mu_j}\frac{\log n}n\right]^{1/2}\right).
\]

The last inequality in the Theorem \ref{th:control} comes from equating
\(e^{-x}\) to \(n^{-c}\) which implies that \(x=c \ln n\).

\hypertarget{control-of-r_1varepsilon2}{%
\subsubsection{\texorpdfstring{Control of
\(\|R_1\varepsilon\|^2\)}{Control of \textbackslash\textbar R\_1\textbackslash varepsilon\textbackslash\textbar\^{}2}}\label{control-of-r_1varepsilon2}}

One can check that \[
\begin{aligned}
R_1\,R_1^T &= H\textrm{diag}(\sqrt{s_i/\hat s_i}-1)\textrm{diag}(\omega_i^2)\textrm{diag}(\sqrt{s_i/\hat s_i}-1)H^T=H\textrm{diag}(\sqrt{s_i/\hat s_i}-1)^2\textrm{diag}(\omega_i^2)H^T \\
&\leq \max_i(\sqrt{s_i/\hat s_i}-1)^2 \Sigma_S^2 J_n.
\end{aligned}
\] As \(J_n\) has only one non-zero eigenvalue which is also its trace
and which is equal to \(n\), it follows using Gendre
(\protect\hyperlink{ref-gendre_model_2014}{2014} Lemma 8.2) that for any
\(x>0\), with probability larger than \(1-e^{-x}\),
\begin{equation}\label{eq:W-gendre}
\frac{\|R_1\varepsilon\|^2}{n\max_i(\sqrt{s_i/\hat s_i}-1)^2\Sigma_S^2} \leq 1+2\sqrt{x}+2x
\end{equation} and \[
\frac{\|R_1\varepsilon\|^2}{\Sigma_S^2} \leq {2(1+2\sqrt{x}+2x)(1+c)(1+o(1))}\frac{1+s_{\max}\bar\rho_S}{{\sum_{j\in S}\mu_j}}\log n
\]

\hypertarget{control-of-r_22}{%
\subsubsection{\texorpdfstring{Control of
\(\|R_2\|^2\)}{Control of \textbackslash\textbar R\_2\textbackslash\textbar\^{}2}}\label{control-of-r_22}}

Simple algebra computations show that \[ 
\begin{aligned}
\|R_2\|^2 &= n\left(\frac {\sqrt{\mu_j^A}} {n_A}\sum_{i\in A} (\sqrt{s_i/\hat s_i}-1) - \frac {\sqrt{\mu_j^B}} {n_B}\sum_{i\in B} (\sqrt{s_i/\hat s_i}-1)\right)^2\\
&\leq n(\sqrt{\mu_j^A}+\sqrt{\mu_j^B})^2 \left(\frac {1} {n_A}\sum_{i\in A} (\sqrt{s_i/\hat s_i}-1) + \frac {1} {n_B}\sum_{i\in B} (\sqrt{s_i/\hat s_i}-1)\right)^2
\end{aligned}
\] such that, using Cauchy-Schwarz inequality and
\(\sqrt{s_i/\hat s_i}-1=\sqrt{s_i}(\sqrt{s_i/\hat s_i}-1)/\sqrt{s_i}\),
we have:
\[ \frac{\|R_2\|}{\sqrt n(\sqrt{\mu_j^A}+\sqrt{\mu_j^B})} \leq \left(\frac {1} {n_A^2}\sum_{i\in A} \frac 1{s_i} + \frac {1} {n_B^2}\sum_{i\in B} \frac 1{s_i}\right)^{1/2} \left(\frac {1} {n_A}\sum_{i\in A} s_i(\sqrt{s_i/\hat s_i}-1)^2 + \frac {1} {n_B}\sum_{i\in B} s_i(\sqrt{s_i/\hat s_i}-1)^2\right)^{1/2}.\]
It follows that \[
\begin{aligned}
\frac{\|R_2\|}{\sqrt n(\sqrt{\mu_j^A}+\sqrt{\mu_j^B})} &\leq \Sigma_S \max_i\left|\sqrt{s_i/\hat s_i}-1\right|\left( \frac {1} {n_A}\sum_{i\in A} s_i+\frac {1} {n_B}\sum_{i\in B} s_i\right)^{1/2}\\
&\leq \Sigma_S \max_i\left|\sqrt{s_i/\hat s_i}-1\right| \sqrt{\max(n/n_A,n/n_B)}.
\end{aligned}
\] Finally \[
\frac{\|R_2\|^2}{\Sigma_S^2}\leq 2(1+c)(1+o(1))\left(\sqrt{\mu_j^A}+\sqrt{\mu_j^B}\right)^2 \times\frac{1+s_{\max}\bar\rho_S}{{\sum_{j\in S}\mu_j}}(\frac n{n_A}\vee\frac{n}{n_B})\,\log n.
\]

\hypertarget{proof-of-lemma}{%
\subsection{\texorpdfstring{Proof of Lemma
\ref{lem:ineqNB}}{Proof of Lemma }}\label{proof-of-lemma}}

\label{proof:lemma}

We recall that \(\hat s_i=nX_{i\bullet}^S\Big/X_{\bullet\bullet}^S\)
where \(X_{i\bullet}^S=\sum_{j\in S}X_{ij}\) and
\(X_{\bullet\bullet}^S=\sum_{i=1}^n\sum_{j\in S} X_{ij}\). Under the
Gaussian approximation, taking into account that \(\sum_{i=1}^n s_i=n\),
we have \[
X_{i\bullet}^S\approx \mathcal N(s_i\,\sum_{j\in S}\mu_j,s_i\,\sum_{j\in S}\mu_j+s_i^2\sum_{j\in S}{\mu_j\rho_j}) \quad \text{and}\quad 
X_{\bullet\bullet}^S\approx \mathcal N(n\,\sum_{j\in S}\mu_j,n\,\sum_{j\in S}\mu_j+\sum_{i=1}^n s_i^2\sum_{j\in S}{\mu_j\rho_j}).
\]

Using Hoeffding's inequality applied to the two latter random variables,
for all \(t>0\), the following inequalities hold with probability larger
than \(1-4\exp(-t^2/2)\): \[
\frac{n\sum_{j\in S}\mu_j-t\,\sqrt{n\sum_{j\in S}\mu_j+\sum_{i=1}^{n}s_i^2\sum_{j\in S}{\mu_j\rho_j}}}{s_i\,\sum_{j\in S}\mu_j+t\,\sqrt{ s_i\,\sum_{j\in S}\mu_j+s_i^2\sum_{j\in S}{\mu_j\rho_j}}}\leq \frac{X_{\bullet\bullet}^S}{X_{i\bullet}^S} \leq \frac{n\sum_{j\in S}\mu_j+t\,\sqrt{n\sum_{j\in S}\mu_j+\sum_{i=1}^n s_i^2\sum_{j\in S}{\mu_j\rho_j}}}{ s_i\,\sum_{j\in S}\mu_j-t\,\sqrt{s_i\,\sum_{j\in S}\mu_j+s_i^2\sum_{j\in S}{\mu_j\rho_j}}}
\] Considering that
\(\sum_{i=1}^n s_i^2\leq s_{\max}\sum_{i=1}^n s_i=n\,s_{\max}\) with
\(s_{\max}=\max(s_i)\), we have \[
\frac{1}{s_i}\frac{\sum_{j\in S}\mu_j-t\,\sqrt{\sum_{j\in S}\mu_j+s_{\max}\sum_{j\in S}{\mu_j\rho_j}}\Big/\sqrt{n}}{\sum_{j\in S}\mu_j+t\,\sqrt{ \sum_{j\in S}\mu_j+s_i\sum_{j\in S}{\mu_j\rho_j}}\Big/\sqrt{s_i}}
\leq \frac{X_{\bullet\bullet}^S}{n X_{i\bullet}^S} \leq 
\frac{1}{s_i}\frac{\sum_{j\in S}\mu_j+t\,\sqrt{\sum_{j\in S}\mu_j+s_{\max}\sum_{j\in S}{\mu_j\rho_j}}\Big/\sqrt{n}}{\sum_{j\in S}\mu_j-t\,\sqrt{\sum_{j\in S}\mu_j+s_i\sum_{j\in S}{\mu_j\rho_j}}\Big/\sqrt{s_i}}
\] Denoting
\[\bar\rho_S:=\sum_{j\in S}\mu_j\rho_j\big/\sum_{j\in S}\mu_j\] we have
as \(s_i\leq n\) \[
\frac{1-t\,\sqrt{1+s_{\max}\,\bar\rho_S}\Big/\sqrt{n\sum_{j\in S}\mu_j}}{1+t\,\sqrt{ 1+s_{\max}\,\bar\rho_S}\Big/\sqrt{n\sum_{j\in S}\mu_j}}
\leq \frac{s_i}{\hat s_i} \leq 
\frac{1+t\,\sqrt{1+s_{\max}\,\bar\rho_S}\Big/\sqrt{n\sum_{j\in S}\mu_j}}{1-t\,\sqrt{ 1+s_{\max}\,\bar\rho_S}\Big/\sqrt{n\sum_{j\in S}\mu_j}}
\]

Using that \((1+x)^{-1}\approx 1-x(1+o(1))\) when \(x\) is small, we get
our inequality\\
\[
\left|\sqrt{\frac{s_i}{\hat s_i}}-1\right|  \leq (1+o(1))t\sqrt{\frac{1+s_{\max}\bar\rho_S}{n\sum_{j\in S}\mu_j}}
\] for \(t^2\) being small in front of
\(n\sum_{i\in S}\mu_j\big/(1+s_{\max}\,\bar\rho_S)\) in order to have
the approximation valid.

\hfill\(\square\)\bigskip

\paragraph*{Acknowledgements}

We are grateful to the Animal Platform UMS 3612 CNRS - US25 Inserm -
Faculté de Pharmacie de Paris, Université Paris Descartes, Paris,
France. We are also grateful to Dr.~Christine Charrueau for having
contributed to the realization of the transcriptomic analysis of the
murine samples.

\paragraph*{Grants}

This work was supported by the Agence Nationale de la Recherche (ANR,
ANR Fibrother ANR-18-CE18-0005-01 to C.H.)

\paragraph*{Animal experiments authorization}

All animal experiments were approved by the Ethics Committee for
Experimentation of Paris Descartes University and authorized by the
French Ministry of National Education, Higher Education and Research
(APAFiS \#4029-2016020513385222v1).

\paragraph*{Data and Code Availability}

Transcriptomic analysis on murine sample together with the code for
simulation will be made available as open data. The code implementing
our procedure will be made available as an open source R software.

\hypertarget{references}{%
\section*{References}\label{references}}
\addcontentsline{toc}{section}{References}

\hypertarget{refs}{}
\leavevmode\hypertarget{ref-anders_differential_2010}{}%
Anders, Simon, and Wolfgang Huber. 2010. ``Differential Expression
Analysis for Sequence Count Data.'' \emph{Genome Biology} 11 (October):
R106. \url{https://doi.org/10.1186/gb-2010-11-10-r106}.

\leavevmode\hypertarget{ref-bullard_evaluation_2010}{}%
Bullard, James H, Elizabeth Purdom, Kasper D Hansen, and Sandrine
Dudoit. 2010. ``Evaluation of Statistical Methods for Normalization and
Differential Expression in mRNA-Seq Experiments.'' \emph{BMC
Bioinformatics} 11 (February): 94.
\url{https://doi.org/10.1186/1471-2105-11-94}.

\leavevmode\hypertarget{ref-chandel_association_2018}{}%
Chandel, Rajeev, Roli Saxena, Ashim Das, and Jyotdeep Kaur. 2018.
``Association of Rno-miR-183-96-182 Cluster with Diethyinitrosamine
Induced Liver Fibrosis in Wistar Rats.'' \emph{Journal of Cellular
Biochemistry} 119 (5): 4072--84.
\url{https://doi.org/https://doi.org/10.1002/jcb.26583}.

\leavevmode\hypertarget{ref-curis_determination_2019}{}%
Curis, Emmanuel, Cindie Courtin, Pierre Alexis Geoffroy, Jean-Louis
Laplanche, Bruno Saubaméa, and Cynthia Marie-Claire. 2019.
``Determination of Sets of Covariating Gene Expression Using Graph
Analysis on Pairwise Expression Ratios.'' Edited by Bonnie Berger.
\emph{Bioinformatics} 35 (2): 258--65.
\url{https://doi.org/10.1093/bioinformatics/bty629}.

\leavevmode\hypertarget{ref-gendre_model_2014}{}%
Gendre, Xavier. 2014. ``Model Selection and Estimation of a Component in
Additive Regression.'' \emph{ESAIM: Probability and Statistics} 18:
77--116. \url{https://doi.org/10.1051/ps/2012028}.

\leavevmode\hypertarget{ref-gregory_mir_200_2008}{}%
Gregory, Philip A., Andrew G. Bert, Emily L. Paterson, Simon C. Barry,
Anna Tsykin, Gelareh Farshid, Mathew A. Vadas, Yeesim Khew-Goodall, and
Gregory J. Goodall. 2008. ``The miR-200 Family and miR-205 Regulate
Epithelial to Mesenchymal Transition by Targeting ZEB1 and SIP1.''
\emph{Nature Cell Biology} 10 (5): 593--601.
\url{https://doi.org/10.1038/ncb1722}.

\leavevmode\hypertarget{ref-hoffmann_hepatic_2020}{}%
Hoffmann, Céline, Nour El Houda Djerir, Anne Danckaert, Julien
Fernandes, Pascal Roux, Christine Charrueau, Anne-Marie Lachagès, et al.
2020. ``Hepatic Stellate Cell Hypertrophy Is Associated with Metabolic
Liver Fibrosis.'' \emph{Scientific Reports} 10 (1): 3850.
\url{https://doi.org/10.1038/s41598-020-60615-0}.

\leavevmode\hypertarget{ref-holm_1979}{}%
Holm, Sture. 1979. ``A Simple Sequentially Rejective Multiple Test
Procedure.'' \emph{Scandinavian Journal of Statistics} 6 (2): 65--70.
\url{http://www.jstor.org/stable/4615733}.

\leavevmode\hypertarget{ref-hu_role_2015}{}%
Hu, Jiangfeng, Chao Chen, Qidong Liu, Baohai Liu, Chenlin Song, Songchen
Zhu, Chaoqun Wu, et al. 2015. ``The Role of the miR-31/FIH1 Pathway in
TGF-\(\beta\)-Induced Liver Fibrosis.'' \emph{Clinical Science (London,
England: 1979)} 129 (4): 305--17.
\url{https://doi.org/10.1042/CS20140012}.

\leavevmode\hypertarget{ref-jiang_roles_2017}{}%
Jiang, Xue-Ping, Wen-Bing Ai, Lin-Yan Wan, Yan-Qiong Zhang, and
Jiang-Feng Wu. 2017. ``The Roles of microRNA Families in Hepatic
Fibrosis.'' \emph{Cell \& Bioscience} 7 (July).
\url{https://doi.org/10.1186/s13578-017-0161-7}.

\leavevmode\hypertarget{ref-karakatsanis_expression_2013}{}%
Karakatsanis, Andreas, Ioannis Papaconstantinou, Maria Gazouli, Anna
Lyberopoulou, George Polymeneas, and Dionysios Voros. 2013. ``Expression
of microRNAs, miR-21, miR-31, miR-122, miR-145, miR-146a, miR-200c,
miR-221, miR-222, and miR-223 in Patients with Hepatocellular Carcinoma
or Intrahepatic Cholangiocarcinoma and Its Prognostic Significance.''
\emph{Molecular Carcinogenesis} 52 (4): 297--303.
\url{https://doi.org/https://doi.org/10.1002/mc.21864}.

\leavevmode\hypertarget{ref-leung_wnt-catenin_2015}{}%
Leung, Wilson K. C., Mian He, Anthony W. H. Chan, Priscilla T. Y. Law,
and Nathalie Wong. 2015. ``Wnt/\(\beta\)-Catenin Activates
MiR-183/96/182 Expression in Hepatocellular Carcinoma That Promotes Cell
Invasion.'' \emph{Cancer Letters} 362 (1): 97--105.
\url{https://doi.org/10.1016/j.canlet.2015.03.023}.

\leavevmode\hypertarget{ref-li_normalization_2012}{}%
Li, J., D. M. Witten, I. M. Johnstone, and R. Tibshirani. 2012.
``Normalization, Testing, and False Discovery Rate Estimation for
RNA-Sequencing Data.'' \emph{Biostatistics} 13 (3): 523--38.
\url{https://doi.org/10.1093/biostatistics/kxr031}.

\leavevmode\hypertarget{ref-li_microrna-34a_2015}{}%
Li, Xiaofei, Yongxin Chen, Shuang Wu, Jinke He, Lianqing Lou, Weiwei Ye,
and Jinhe Wang. 2015. ``microRNA-34a and microRNA-34c Promote the
Activation of Human Hepatic Stellate Cells by Targeting Peroxisome
Proliferator-Activated Receptor \(\gamma\).'' \emph{Molecular Medicine
Reports} 11 (2): 1017--24. \url{https://doi.org/10.3892/mmr.2014.2846}.

\leavevmode\hypertarget{ref-DESeq2}{}%
Love, Michael I., Wolfgang Huber, and Simon Anders. 2014. ``Moderated
Estimation of Fold Change and Dispersion for Rna-Seq Data with Deseq2.''
\emph{Genome Biology} 15 (12): 550.
\url{https://doi.org/10.1186/s13059-014-0550-8}.

\leavevmode\hypertarget{ref-marioni_rna-seq:_2008}{}%
Marioni, John C., Christopher E. Mason, Shrikant M. Mane, Matthew
Stephens, and Yoav Gilad. 2008. ``RNA-Seq: An Assessment of Technical
Reproducibility and Comparison with Gene Expression Arrays.''
\emph{Genome Research} 18 (9): 1509--17.
\url{https://doi.org/10.1101/gr.079558.108}.

\leavevmode\hypertarget{ref-massart_tight_1990}{}%
Massart, P. 1990. ``The Tight Constant in the Dvoretzky-Kiefer-Wolfowitz
Inequality.'' \emph{The Annals of Probability} 18 (3): 1269--83.
\url{http://www.jstor.org/stable/2244426}.

\leavevmode\hypertarget{ref-mortazavi_mapping_2008}{}%
Mortazavi, Ali, Brian A. Williams, Kenneth McCue, Lorian Schaeffer, and
Barbara Wold. 2008. ``Mapping and Quantifying Mammalian Transcriptomes
by RNA-Seq.'' \emph{Nature Methods} 5 (7): 621--28.
\url{https://doi.org/10.1038/nmeth.1226}.

\leavevmode\hypertarget{ref-murakami_progression_2011}{}%
Murakami, Yoshiki, Hidenori Toyoda, Masami Tanaka, Masahiko Kuroda,
Yoshinori Harada, Fumihiko Matsuda, Atsushi Tajima, Nobuyoshi Kosaka,
Takahiro Ochiya, and Kunitada Shimotohno. 2011. ``The Progression of
Liver Fibrosis Is Related with Overexpression of the miR-199 and 200
Families.'' \emph{PLOS ONE} 6 (1): e16081.
\url{https://doi.org/10.1371/journal.pone.0016081}.

\leavevmode\hypertarget{ref-edgeR}{}%
Robinson, Mark D, Davis J McCarthy, and Gordon K Smyth. 2010. ``EdgeR: A
Bioconductor Package for Differential Expression Analysis of Digital
Gene Expression Data.'' \emph{Bioinformatics} 26 (1): 139--40.

\leavevmode\hypertarget{ref-robinson_scaling_2010}{}%
Robinson, Mark D., and Alicia Oshlack. 2010. ``A Scaling Normalization
Method for Differential Expression Analysis of RNA-Seq Data.''
\emph{Genome Biology} 11 (March): R25.
\url{https://doi.org/10.1186/gb-2010-11-3-r25}.

\leavevmode\hypertarget{ref-tessitore_microrna_2016}{}%
Tessitore, Alessandra, Germana Cicciarelli, Filippo Del Vecchio, Agata
Gaggiano, Daniela Verzella, Mariafausta Fischietti, Valentina
Mastroiaco, et al. 2016. ``MicroRNA Expression Analysis in High Fat
Diet-Induced NAFLD-NASH-HCC Progression: Study on C57BL/6J Mice.''
\emph{BMC Cancer} 16 (January).
\url{https://doi.org/10.1186/s12885-015-2007-1}.

\leavevmode\hypertarget{ref-vandesompele_accurate_2002}{}%
Vandesompele, Jo, Katleen De Preter, Filip Pattyn, Bruce Poppe, Nadine
Van Roy, Anne De Paepe, and Frank Speleman. 2002. ``Accurate
Normalization of Real-Time Quantitative RT-PCR Data by Geometric
Averaging of Multiple Internal Control Genes.'' \emph{Genome Biology} 3
(7): research0034.1--research0034.11.
\url{https://www.ncbi.nlm.nih.gov/pmc/articles/PMC126239/}.

\leavevmode\hypertarget{ref-yang_microrna-802_2019}{}%
Yang, Xi, Hanying Xing, Jingzhen Liu, Linquan Yang, Huan Ma, and Huijuan
Ma. 2019. ``MicroRNA-802 Increases Hepatic Oxidative Stress and Induces
Insulin Resistance in High-Fat Fed Mice.'' \emph{Molecular Medicine
Reports} 20 (2): 1230--40. \url{https://doi.org/10.3892/mmr.2019.10347}.

\leavevmode\hypertarget{ref-zhang_mir-582-5p_2015}{}%
Zhang, Yi, Wei Huang, Yan Ran, Yan Xiong, Zibiao Zhong, Xiaoli Fan,
Zhenghua Wang, and Qifa Ye. 2015. ``miR-582-5p Inhibits Proliferation of
Hepatocellular Carcinoma by Targeting CDK1 and AKT3.'' \emph{Tumor
Biology} 36 (11): 8309--16.
\url{https://doi.org/10.1007/s13277-015-3582-0}.

\leavevmode\hypertarget{ref-zhen_microrna-802_2018}{}%
Zhen, Yun-Feng, Yun-Jia Zhang, Hang Zhao, Hui-Juan Ma, and Guang-Yao
Song. 2018. ``MicroRNA-802 Regulates Hepatic Insulin Sensitivity and
Glucose Metabolism.'' \emph{International Journal of Clinical and
Experimental Pathology} 11 (5): 2440--9.
\url{https://www.ncbi.nlm.nih.gov/pmc/articles/PMC6958293/}.

\end{document}